\documentclass{birkjour}
 \theoremstyle{definition}
 
 \theoremstyle{remark}

 \numberwithin{equation}{section}

\begin{document}

%
%
%
%
%
%
%
%
%

\title[Super Quantum Mechanics in the Integral Form Formalism]
 {Super Quantum Mechanics in the Integral Form Formalism}

\author[L. Castellani]{L. Castellani}

\address{%
Dipartimento di Scienze e Innovazione Tecnologica, \\
Universit\`a del Piemonte Orientale,
Viale T. Michel, 11, 15121 Alessandria, Italy\\
INFN, Sezione di Torino, via P. Giuria 1, 10125 Torino, Italy\\
Arnold-Regge Center, via P. Giuria 1,  10125 Torino, Italy
}
\email{leonardo.castellani@uniupo.it}

\author{R. Catenacci}
\address{
Dipartimento di Scienze e Innovazione Tecnologica, \\
Universit\`a del Piemonte Orientale,
Viale T. Michel, 11, 15121 Alessandria, Italy\\
Arnold-Regge Center, via P. Giuria 1,  10125 Torino, Italy \\
Gruppo Nazionale di Fisica Matematica, INdAM, P.le Aldo Moro 5, 00185 Roma, Italy
}
\email{roberto.catenacci@uniupo.it}

\author{P.A. Grassi}
\address{
Dipartimento di Scienze e Innovazione Tecnologica, \\
Universit\`a del Piemonte Orientale,
Viale T. Michel, 11, 15121 Alessandria, Italy\\
INFN, Sezione di Torino, via P. Giuria 1, 10125 Torino, Italy\\
Arnold-Regge Center, via P. Giuria 1,  10125 Torino, Italy \\
Center for Gravitational Physics, Yukawa Institute for Theoretical Physics, \\
Kyoto University, Kyoto 606-8502, Japan
}
\email{pietro.grassi@uniupo.it}



\date{July 11, 2017}

\begin{abstract}
We reformulate Super Quantum Mechanics
in the context of integral forms. This framework allows to interpolate between different
actions for the same theory, connected by different choices of Picture Changing
Operators (PCO).  In this way we retrieve component and superspace actions, and prove their equivalence.
The PCO are closed integral forms, and can be interpreted as super Poincar\'e duals of bosonic submanifolds embedded
into a supermanifold.. We use them to construct Lagrangians that are top integral forms, and therefore can be integrated on the whole supermanifold.  The $D=1, ~N=1$ and the $D=1,~ N=2$ cases are studied, in a flat and in a curved supermanifold. In this formalism we also consider coupling with gauge fields, Hilbert space of quantum states and observables.
\end{abstract}

\maketitle

\section{Introduction}

Since the invention of supersymmetry, several authors provided useful
mathematical tools for its geometrical formulation, based essentially on the
interpretation of supersymmetry as a coordinate transformation in fermionic
directions, described by Grassmann coordinates $\theta$. Still, there
remained several problems, mostly related to integration theory on
supermanifolds.

The first formulations of supersymmetric models were given in terms of a
\textit{component} action, containing bosonic and fermionic fields, and
invariant under supersymmetry transformations mixing bosons and fermions.

The same dynamics can be derived in a more efficient way from an action
which is \textsl{manifestly} invariant under supersymmetry. This framework
is known as \textit{superspace} approach and the various fields of the
spectrum are contained in some \textit{superfields} (or superforms). The
action is obtained as an integral of products of superfields and their
derivatives. In this approach the set of coordinates $x$ of the worldvolume
is augmented by a set of fermionic coordinates $\theta $, and a superfield
is a function of $x$ and $\theta $. These coordinates parametrize an open
set of a supermanifold (which locally is denoted by ${\mathbb{R}}^{(n|m)}$)
which is a generalization of a differential manifold. In the section 2 we
summarize the theory of supermanifolds, on which a vast literature exists
(see for ex. \cite%
{Varadarajan:2004yz,Bruzzo,Freed,Fioresi:2007zz,Catenacci:2010cs}).
Fermionic derivatives are needed in this context, and they form an algebra
representing the supersymmetry algebra. A supersymmetry variation of a
superfield is obtained by means of a differential operator, representing the
supersymmetry generators on the ring of superfunctions on the supermanifold.
In this framework, the action is manifestly supersymmetric since the
supersymmetry variation of the Lagrangian is a total derivative on
superspace, and its integral vanishes. Although the superspace framework has
several advantages w.r.t. to the component formalism, the geometry behind it
still needs some clarification.

Motivated by string theory (both in RNS formalism \cite%
{Witten:2012bg,Witten:2012bh} and in Pure Spinor formalism \cite%
{Berkovits:2004px}) new geometrical elements, known as integral forms, were
introduced. They are essential to provide a sensible theory of geometric
integration for supermanifolds and they are the natural generalizations of
differential forms of a conventional manifold. Their properties and their
integration theory are briefly described in the text and we refer to the
literature \cite{Belo,Witten:2012bg,LMP,Castellani:2015paa} and the book by
Voronov \cite{voronov-book}) for more details.

Once integration on supermanifolds has been established on a sound
geometrical basis, we can finally rewrite the action in the component
formalism and the action in superspace as different representations of the
same geometrical action. This is achieved by constructing an interpolating
action, known in the literature as a \textit{rheonomic} action (see the main
reference \cite{cube}). The Picture Changing Operators can be interpreted as
integral forms ${\mathbb{Y}}$ representing the super Poincar\'e dual of the
embedding of a bosonic submanifold into a supermanifold, and are used to
construct a Lagrangian (a $n$-superform multiplied by a PCO ${\mathbb{Y}}$
to give a top integral form) that can be integrated on the whole
supermanifold. It turns out that by choosing different Poincar\'e duals ${%
\mathbb{Y}}$ one can interpolate between different equivalent actions.

To illustrate these features, we consider in this paper the simple example
of Super Quantum Mechanics, viewed as a $D=1$ quantum field theory. The
application of the formalism of integral forms to theories in higher
dimensions will be the subject of a forthcoming paper. The case of $D=3$ $N=1
$ supergravity was analyzed in \cite{3dsuper}). We consider both $N=1$ and
the $N=2$ cases, since they have different characteristics worth to be
described. First, we build the rheonomic action (which was not present in
the literature), then we show how the different choices of ${\mathbb{Y}}$
interpolate between the different realisations (component action or
superspace action). In the case of SQM everything is clear and easy to
compute and provides a perfect introductory example for the use of these
techniques. In the last section, we also argue that the observables of the
theory share the same properties of the action and that also for them one
can use different representations corresponding to suitable ${\mathbb{Y}}$.

The paper is structured as follows: in sec.2 we collect some introductory
material about supermanifold theory and we give also a few mathematical
details about the super particle model. This section can be skipped by experts 
on supermanifold theory. In sec.3, we review the most
important points of the integration theory of integral forms. In sec.4 we
discuss some of the properties of the Picture Changing Operators. In sec.5
we study the model of SQM $N=1$ and in sec.6 the model $N=2$. In sec.7
we discuss the Hilbert space and in sec.8 the observables.

\section{Supermanifolds and the Supersymmetric point particle.}

We give in this introductory section a very short review of the definitions
and the concepts of the theory of supermanifolds. The definitions and the
notations are mainly taken from \cite{Varadarajan:2004yz} and \cite{Bruzzo}
to which we refer for a more complete treatment.

This section also contains some comments and some examples that might help
to gain intuition on the topic.

A \textbf{supercommutative} ring is a $\mathbb{Z}_{2}$-graded ring $%
A=A_{0}\oplus A_{1}$ such that if $i,j\in\mathbb{Z}_{2}$, then $%
a_{i}a_{j}\in A_{i+j}$ and $a_{i}a_{j}=(-1)^{i+j}a_{j}a_{i}$, where $%
a_{k}\in A_{k}$. Elements in $A_{0}$ (resp. $A_{1}$) are called \textbf{even}
(resp. \textbf{odd}).

A \textbf{super ringed space} is a topological space $X$ together with a
sheaf $\mathcal{O}_{X}$ of supercommutative rings. If \ the stalks are local%
\footnote{%
A ring\ is called local if it has a unique maximal ideal.} rings, the super
ringed space is called a \textbf{superspace}.

A \textbf{superdomain} $U^{n|m}$ is the superspace $\left( U^{n},\mathcal{C}%
^{\infty n|m}\right) $, where $U^{n}\subseteq\mathbb{R}^{n}$ is open and $%
\mathcal{C}^{\infty n|m}$ is the sheaf of supercommutative rings given by:
\[
V\mapsto\mathcal{C}^{\infty}\left( V\right) \left[ \theta^{1},\theta
^{2},...,\theta^{m}\right] ,
\]
where $V\subseteq U^{n}$ is open and $\theta^{1},\theta^{2},...,\theta^{m}$
are the generators of a Grassmann algebra. The grading is the natural $%
\mathbb{Z}_{2}$ grading in even and odd elements.

Every element of $\mathcal{C}^{\infty n|m}\left( V\right) $ is called a
\textbf{superfunction} and may be written as $\sum_{I}f_{I}\theta^{I}$,
where $I$ is a multi-index and $f_{I}\in\mathcal{C}^{\infty}\left( V\right) $
is an ordinary function.

A smooth \textbf{supermanifold} $\mathcal{M}$ of dimension $n|m$ is a
superspace locally isomorphic to the superspace $U^{n|m}$. In this section
we will denote by $\mathcal{M}\equiv\left( M,\mathcal{O}_{M}\right) $ the
supermanifold whose underlying topological space is $M$ and whose sheaf of
supercommutative rings is $\mathcal{O}_{M}$ . In the following sections the
supermanifold will be denoted by $\mathcal{M}^{(n|m)}$.

This definition means that given a point $x\in M$ (the underlying
topological space) there is an open set $U\subset M$ such that $U$ is
homeomorphic to $U_{0}\subset\mathbb{R}^{n}$ and $\mathcal{O}_{M}|_{U}$ is
isomorphic to $\mathcal{C}^{\infty n|m}|_{U_{0}}.$ The coordinates $x^{i}$
of $U^{n}$ are called even (or bosonic) coordinates, while the elements $%
\theta^{\alpha}$ are called odd (or fermionic) coordinates.

This definition has a difficulty that arises because, in order to use
supermanifolds in physical applications, we would like to think in terms of
points and functions, but ordinary (topological) points here are only the
points of the topological space $M$ , and the (super)functions are really
sections of the sheaf.

To a section $s$ of $\mathcal{O}_{M}$ on an open set $U$ of $M$ containing $%
x:$
\[
s:U\rightarrow\mathcal{O}_{M}|_{U}
\]
one can associate its \textbf{value} in $x$ as the unique real number $%
s^{\sim}\left( x\right) $ such that $s-s^{\sim}\left( x\right) $ is not
invertible on every neighborhood of $x$ contained in $U$.

The sheaf of algebras $\mathcal{O}^{\sim}$, whose sections are the functions
$s^{\sim}$, defines the structural sheaf of an ordinary differentiable
manifold on the space $M$. This manifold is called the \textbf{reduced
manifold} $\mathcal{M}_{\mathrm{red}}$ of the supermanifold $\mathcal{M}$.
The \textbf{points} of $\mathcal{M}$ are just the ordinary points of $%
\mathcal{M}_{\mathrm{red}}.$

For example, in the simple case of the supermanifold $\mathbb{R}%
^{1|1}=\left( \mathbb{R},\mathcal{C}^{\infty}\left( \mathbb{R}\right) \left[
\theta\right] \right) $, the points are the ordinary points of $\mathbb{R}.$
The global section $s=x\theta$ is nilpotent and for any real number $a\neq0$%
, $x\theta-a$ is invertible (its inverse being $-a^{-2}x\theta-a^{-1}$) and
hence the value of $s$ is \textbf{zero} at any point $x\in\mathbb{R}_{%
\mathrm{red}}^{1|1}=\mathbb{R}.$ In other terms the \textbf{value} at the
point $x$ of a generic section $\sigma=f+g\theta$ is simply $f(x),$ the
value in $x$ of the ordinary real function $f.$ This means that the
topological points cannot see the nilpotent objects because we cannot
reconstruct a section from its values at the topological points, and this is
not what is needed to support the intuition in physical applications,
because we would like to give a meaning to odd functions.

We can consider also the case of the purely odd supermanifold $\mathbb{R}%
^{0|1}=\left( \mathbb{R}^{0},\mathcal{C}^{\infty}\left( \mathbb{R}%
^{0}\right) \left[ \theta\right] \right) .$ The ring of "smooth" functions
is simply $\mathbb{R}\left[ \theta\right] /\theta^{2}.$ The reduced manifold
is just a single point\footnote{%
The ring $\mathbb{R}\left[ \theta\right] /\theta^{2}$ has only one prime
(and maximal) ideal (the ideal generated by $\theta$ ) and hence we verify
that the set of points of $\mathbb{R}^{0|1}$ is the spectrum of the ring of
smooth functions.}. If we want to analyze "geometrically" $\mathbb{R}^{0|1}$
we must study the maps from $\mathbb{R}^{0|1}$ to an ordinary manifold $M.$
These maps can be represented by ring omomorphisms going in the opposite
direction:%
\[
\mathcal{C}^{\infty}\left( M\right) \rightarrow\mathbb{R}\left[ \theta\right]
/\theta^{2}
\]
 A generic homomorphism is given by $f\rightarrow A(f)+B(f)\theta.$ This
gives two equations, the first one is $A(fg)=A(f)A(g)$ from which we
conclude that $A(f)=f(m)$ with $m$ a point of $M.$ The second one is $%
B(fg)=B(f)g(m)+B(g)f(m)$ which states that $B(f)$ is a derivation over
functions and hence is given by a tangent vector at $m$: $B(f)=$ $\xi_{m}(f).
$ We can describe $\mathbb{R}^{0|1}$ in $M$ as a point with a set of tangent
vectors or, more abstractly, as a "nilpotent cloud" surrounding a single
abstract point.

Maps in the opposite direction, from an ordinary manifold $M$ to $\mathbb{R}%
^{0|1}$ are given by homomorphisms:%
\[
\mathbb{R}\left[ \theta\right] /\theta^{2}\rightarrow\mathcal{C}^{\infty
}\left( M\right)
\]
We have that $\theta$ must go to zero (because $\mathcal{C}^{\infty}\left(
M\right) $ has no nilpotents) and hence any such map simply maps the
manifold $M$ to the single point in $\mathbb{R}^{0|1}$ and we see again that
we cannot give a meaning to odd functions. The same argument holds true also
in the general case of $\mathbb{R}^{0|m}.$

This problem can be solved using the idea of the \textbf{functor of points}
which is the formalization of the concept of \textit{auxiliary fermionic
parameters} often used in physical applications.

This functor can be used for giving a definite meaning to the elusive
concept of odd functions i.e. "classical fermions" (following the
terminology of \cite{Freed})

We now wish to explain how the intuitive geometrical interpretation of the $%
x^{i}$'s as \textquotedblleft even coordinates\textquotedblright\ and the $%
\theta^{\alpha}$'s as \textquotedblleft odd coordinates\textquotedblright\
can be obtained from the super ringed space definition of supermanifolds
through the concept of \textbf{functor of points}.

Given two supermanifolds $\mathcal{M}$ and $\mathcal{S}$, the $\mathcal{S}$%
-points of $\mathcal{M}$ (or the points of $\mathcal{M}$ parametrized by $%
\mathcal{S}$) are given by the set
\[
\mathcal{M}(\mathcal{S})=\mathrm{Hom}(\mathcal{S},\mathcal{M})=\{\mathrm{%
set\,\,of\,\,morphisms\,}\mathcal{\,S}\rightarrow\mathcal{M}\}\mathrm{\,.}
\]
$\mathcal{M}$ is the supermanifold we want to describe and $\mathcal{S}$ is
the model on which we base the description of $\mathcal{M}$. Changing $%
\mathcal{S}$ modifies the description of $\mathcal{M}$. The functor which
associates $\mathcal{S}$ to $\mathcal{M}(\mathcal{S})$ is a functor between
the category of supermanifolds and the category of sets.

The set of morphisms $\mathcal{S}\rightarrow\mathcal{M}$ is, in this
construction, the set of \textquotedblleft points\textquotedblright\ of the
supermanifolds. See also \cite{Fioresi:2007zz} for more details.

Let us recall now some fundamental properties of morphisms. A morphism $f$
between two superdomains $U^{p|q}$ and $V^{r|s}$ is given by a smooth map $%
f^{\sim}:U^{p}\rightarrow V^{r}$ and a homomorphism $f^{\ast}$ of
superalgebras that respects the parity:
\[
f^{\ast}:\mathcal{C}^{\infty\,r|s}(V^{r})\rightarrow\mathcal{C}^{\infty
\,p|q}(U^{p}).
\]
It must satisfy the following properties:

\begin{itemize}
\item If $t=(x^{1},\ldots,x^{r})$ are coordinates on $V^{r}$, each component
$x^{i}$ can also be interpreted as a section of $\mathcal{C}^{\infty
\,r|s}(V^{r})$. If $f^{i}=f^{\ast}(x^{i})$, then $f^{i}$ is an \textbf{even}
element of the algebra $\mathcal{C}^{\infty\,p|q}(U^{p})$.

\item If $\theta^{\alpha}$ is a generator of $\mathcal{C}^{\infty%
\,r|s}(V^{r})$, then $g^{\alpha}=f^{\ast}(\theta^{\alpha})$ is an \textbf{odd%
} element of the algebra $\mathcal{C}^{\infty\,p|q}(U^{p})$.

\item The smooth map $f^{\sim}:U^{p}\rightarrow V^{r}$ is $%
f^{\sim}=(f^{1\sim },\ldots,f^{r\sim})$, where the $f^{i\sim}$ are the
values of the even elements $f^{i}$.
\end{itemize}

The following fundamental result (see for example \cite{Varadarajan:2004yz})
gives the local characterizations of morphisms:

{If} $\phi:U^{p}\rightarrow V^{r}$ is a smooth map and $f^{i},g^{\alpha}$,
with $i=1,\ldots,r$, $\alpha=1,\ldots,s$, are given elements of $\mathcal{C}%
^{\infty\,p|q}(U^{p})$, with $f^{i}$ even, $g^{\alpha}$ odd, and satisfying $%
\phi=(f^{1\sim},\ldots,f^{r\sim})$, there exists a unique morphism $%
f:U^{p|q}\rightarrow V^{r|s}$ with $f^{\sim}=\phi$ , $f^{\ast}(x^{i})=f^{i}$
and $f^{\ast}(\theta^{\alpha})=g^{\alpha}$.

A morphism $f\in\mathrm{Hom}(U^{p|q},V^{r|s})$ is then uniquely determined
by a choice of $r$ even sections and $s$ odd sections of $%
C^{\infty\,p|q}(U^{p})$, i.e. morphisms are in one to one correspondence
with $(r+s)$-tuples $(f^{1},\ldots,f^{r},g^{1},\ldots,g^{s})$, where the $%
f^{i}$'s are even and the $g^{\alpha}$'s are odd in the algebra $%
C^{\infty\,p|q}(U^{p})$. If we denote by $\Gamma_{q}^{0}(U^{p})$ and $%
\Gamma_{q}^{1}(U^{p})$ respectively the set of even and odd sections of $%
C^{\infty\,p|q}(U^{p})$, then the above fact is expressed as
\[
\mathrm{Hom}(U^{p|q},V^{r|s})=(\Gamma_{q}^{0}(U^{p|q}))^{r}\times(\Gamma
_{q}^{1}(U^{p|q}))^{s}.
\]
Where $q$ denotes the {}number of odd generators of the algebra we are
considering.

In particular, if $\mathcal{S}=\mathbb{R}^{0|q}$, then
\[
\mathrm{Hom}(\mathbb{R}^{0|q},V^{r|s})=(\Gamma_{q}^{0})^{r}\times(\Gamma
_{q}^{1})^{s}
\]
where $\Gamma_{q}^{0}$ and $\Gamma_{q}^{1}$ represent respectively the even
and the odd components of a Grassmann algebra with $q$ generators, and their
cartesian powers the $r$ "bosonic coordinates" and the $s$ "fermionic
coordinates" of $V^{r|s}$.

One could say that the {}super ringed space structure of $\mathcal{M}$
encodes the information of how the even and odd coordinates $(x^{i},\theta
^{\alpha })$ glue together, but independently on the number of generators of
the underlying super algebra. The number of generators ($q$ in the above
case) can be fixed by taking a supermanifold $\mathcal{S}$ and constructing $%
\mathrm{Hom}(\mathcal{S},\mathcal{M})$. This procedure is the formalization
of the concept of "auxiliary fermionic parameters" often used in physical
applications.

\subsection{The Supersymmetric point particle.}

We now describe the "supersymmetric point particle" as an example of the
general theory. This one-dimensional model is simple with respect to
computations, but it is not at all simple from the mathematical point of
view because the naive interpretation of the supermanifold $\mathbb{R}%
^{(1|1)}$ as a space in which there are "points" with commuting and
anticommuting coordinates $\left( x,\theta \right) $ is not adequate. The
main reason is that in the naive interpretation of $\mathbb{R}^{(1|1)}$
there is only one real coordinate $x$ and only one fermionic coordinate $%
\theta ,\ $so for supersymmetry we are forced to write down equations that
apparently are not allowed or meaningful.

A supersymmetric particle is described by a map $\mathbb{R}\rightarrow V.$
The space $V$ is a real $\mathbb{Z}_{2}$-graded vector space $V=V^{0}\oplus
V^{1}.$ We will consider the simple case in which $V^{0}\oplus V^{1}=\mathbb{%
R\oplus R}^{0|1}.$ The bosonic part is an ordinary function $\varphi (t)\in
\mathbb{R}$, the fermionic part $\nu (t)$ must be defined using the functor
of points.

Let us consider again the example of $\ \mathbb{R}^{0|1}=\left( \mathbb{R}%
^{0},\mathcal{C}^{\infty }\left( \mathbb{R}^{0}\right) \left[ \theta \right]
\right) ;$ we have seen that there are no odd functions from an ordinary
manifold to $\mathbb{R}^{0|1}$ (or $\mathbb{R}^{0|m})$, so we must consider
instead maps from a supermanifold to $\mathbb{R}^{0|1}$. We take\footnote{%
Note that $\mathbb{R}^{1|2}$ is just the minimal choice, we could have taken
$\mathbb{R}^{1|m}$ with $m\geq 2.$} the supermanifold $\mathbb{R}%
^{1|2}=\left( \mathbb{R},\mathcal{C}^{\infty }\left( \mathbb{R}\right) \left[
\eta ^{1},\eta ^{2}\right] \right) $. In terms of the functor of points
description the \textquotedblleft $\mathbb{R}^{1|2}-$ points" of $\mathbb{R}%
^{0|1}$ will be labelled by \textbf{zero} even sections and \textbf{one} odd
section of $C^{\infty }\left( \mathbb{R}^{1|2}\right) $. A map $\mathbb{R}%
^{1|2}\rightarrow \mathbb{R}^{0|1}$ is represented by a morphism
\[
\mathcal{C}^{\infty }\left( \mathbb{R}^{0}\right) \left[ \theta \right]
\rightarrow \mathcal{C}^{\infty }\left( \mathbb{R}\right) \left[ \eta
^{1},\eta ^{2}\right]
\]%
given by: $\theta \rightarrow \sum f_{i}\eta ^{i}$ for some functions $%
f_{i}\in $ $\mathcal{C}^{\infty }\left( \mathbb{R}\right) .$ For any $t\in
\mathbb{R}$ we can consider the family of sections $\nu _{t}$ of $\mathcal{C}%
^{\infty }\left( \mathbb{R}\right) \left[ \eta ^{1},\eta ^{2}\right] $ that
we can now write:%
\begin{equation}
\nu _{t}=\nu (t)=\sum f_{i}(t)\eta ^{i}
\end{equation}%
$\nu (t)$ can be interpreted as an \textit{odd function} that can be used
like an ordinary function and might be called a \textit{classical fermion}.

The Lagrangian of the model is the sum of a \textquotedblleft bosonic part"
and a \textquotedblleft fermionic" one:

\begin{equation}
L=\left( \frac{d\varphi(t)}{dt}\right) ^{2}+\nu(t)\frac{d}{dt}\nu(t)
\label{lagrangiana}
\end{equation}
Note that $\nu^{2}=0$ and hence $\nu(t)\frac{d}{dt}\nu(t)$ is not a total
derivative. This term is bosonic and $\neq0.$

The supersymmetry transformations are defined as:
\begin{equation}
\delta _{\epsilon }\varphi (t)=\epsilon \nu (t)\,,~~~~~\delta _{\epsilon
}\nu (t)=-\epsilon \frac{d\varphi (t)}{dt}\,.
\end{equation}%
Where $\epsilon $ is the constant anticommuting parameter of the
supersymmetry interpreted now as a linear combination of $\eta ^{1}$ and $%
\eta ^{2}$. The Lagrangian transforms as:%
\begin{equation}
\delta _{\epsilon }L=\epsilon \frac{d}{dt}\left( \frac{d\varphi (t)}{dt}\nu
(t)\right)
\end{equation}%
We see that the action $S$ is invariant (with suitable boundary conditions):%
\begin{equation}
\delta _{\epsilon }S=\int_{\mathbb{R}}\delta _{\epsilon }Ldt=0
\end{equation}

\section{Integral forms, integration theory and Poincar\'{e} duals.}

The integral forms are the crucial ingredients to define a geometric
integration theory for supermanifolds inheriting all good properties of
differential forms integration theory in conventional (purely bosonic)
geometry. In this section we briefly describe the notations and the most
relevant definitions (see \cite{Witten:2012bg}, \cite{voronov-book} and also
\cite{Castellani:2015paa,Castellani:2014goa,Castellani:2015ata}).

We consider a supermanifold with $n$ bosonic dimensions and $m$ fermionic
dimensions, denoted here and in the following by $\mathcal{M}^{(n|m)}$ ,
locally isomorphic to the superspace $\mathbb{R}^{(n|m)}$. The local
coordinates in an open set are denoted by $(x^{a},\theta ^{\alpha })$. A $%
(p|q)$ pseudoform $\omega ^{(p|q)}$ has the following structure:
\begin{equation}
\omega ^{(p|q)}=\omega (x,\theta )dx^{a_{1}}\dots dx^{a_{r}}d\theta ^{\alpha
_{1}}\dots d\theta ^{\alpha _{s}}\delta ^{(b_{1})}(d\theta ^{\beta
_{1}})\dots \delta ^{(b_{q})}(d\theta ^{\beta _{q}})  \label{pseudo}
\end{equation}%
where, in a given monomial, the $d\theta ^{a}$ appearing in the product are
different from those appearing in the delta's $\delta (d\theta )$ and $%
\omega (x,\theta )$ is a set of superfields with index structure $\omega
_{\lbrack a_{1}\dots a_{r}](\alpha _{1}\dots \alpha _{s})[\beta _{1}\dots
\beta _{q}]}(x,\theta )$.

The two integer numbers $p$ and $q$ correspond respectively to the \textit{%
form} number and the \textit{picture} number, and they range from $-\infty$
to $+\infty$ for $p$ and $0\leq q\leq m$. The index $b$ on the delta $%
\delta^{(b)}(d\theta^{\alpha})$ denotes the degree of the derivative of the
delta function with respect to its argument. The total picture of $%
\omega^{(p|q)}$ corresponds to the total number of delta functions and its
derivatives. We call $\omega^{(p|q)}$ a \textit{superform} if $q=0$ and an
\textit{integral form} if $q=m$; otherwise it is called \textit{pseudoform}.
The total form degree is given by $p=r+s-\sum_{i=1}^{i=q}b_{i}$ since the
derivatives act effectively as negative forms and the delta functions carry
zero form degree. We recall the following properties:
\begin{equation}
d\theta^{\alpha}\delta(d\theta^{\alpha})=0,\text{ }d\delta^{(b)}(d\theta^{%
\alpha})=0\,,~d\theta^{\alpha}\delta^{(b)}(d\theta^{\alpha
})=-b\delta^{(b-1)}(d\theta^{\alpha})\,,~~b>0\,.~~~
\label{proprietadistrib}
\end{equation}
The index $\alpha$ is not summed. The indices $a_{1}\dots a_{r}$ and $%
\beta_{1}\dots\beta_{q}$ are anti-symmetrized, the indices $%
\alpha_{1}\dots\alpha_{s}$ are symmetrized because of the rules of the
graded wedge product:
\begin{align}
dx^{a}dx^{b} & =-dx^{b}dx^{a}\,,~~~dx^{a}d\theta^{\alpha}=d\theta^{\alpha
}dx^{a}\,,~~~d\theta^{\alpha}d\theta^{\beta}=d\theta^{\beta}d\theta^{\alpha
}\,,  \label{wedgevari} \\
\delta(d\theta^{\alpha})\delta(d\theta^{\beta}) & =-\delta(d\theta^{\beta
})\delta(d\theta^{\alpha})\,, \\
~~~dx^{a}\delta(d\theta^{\alpha}) &=-\delta
(d\theta^{\alpha})dx^{a}\,,~~~d\theta^{\alpha}\delta(d\theta^{\beta})=%
\delta(d\theta^{\beta})d\theta^{\alpha}\,\,.
\end{align}

As usual the module of $(p|q)$ pseudoforms is denoted by $\Omega^{(p|q)}$;
if $q=0$ or $q=m$ it is finitely generated.

It is possible to define the integral over the superspace $\mathbb{R}^{(n|m)}
$ of an \textit{integral top} form $\omega^{(n|m)}$ that can be written
locally as:
\begin{equation}
\omega^{(n|m)}=f(x,\theta)dx^{1}\dots dx^{n}\delta(d\theta^{1})\dots
\delta(d\theta^{m})\,   \label{sezioneberezin}
\end{equation}
where $f(x,\theta)$ is a superfield. By changing the $1$-forms $%
dx^{a},d\theta^{\alpha}$ as $dx^{a}\rightarrow
E^{a}=E_{m}^{a}dx^{m}+E_{\mu}^{a}d\theta^{\mu}$ and $d\theta^{\alpha}%
\rightarrow E^{\alpha}=E_{m}^{\alpha }dx^{m}+E_{\mu}^{\alpha}d\theta^{\mu}$,
we get
\begin{equation}
\omega\rightarrow\mathrm{sdet}(E)\,f(x,\theta)dx^{1}\dots dx^{n}\delta
(d\theta^{1})\dots\delta(d\theta^{m})
\end{equation}
where $\mathrm{sdet}(E)$ is the superdeterminant of the supervielbein $%
(E^{a},E^{a})$.

The integral form $\omega ^{(n|m)}$ can be also viewed as a superfunction $%
\omega (x,\theta ,dx,d\theta )$ on the \textit{odd} dual\footnote{%
In order to make contact with the standard physics literature we adopt the
conventions that $d$ is an odd operator and $dx$ (an odd form) is dual to
the even vector $\frac{\partial }{\partial x}$. The same holds for the even
form $d\theta $ dual to the odd vector $\frac{\partial }{\partial \theta }.$
As clearly explained for example in the appendix of the paper \cite{VORONOV3}
if one introduces also the natural concept of even differential (in order to
make contact with the standard definition of cotangent bundle of a manifold)
our cotangent bundle (that we consider as the bundle of one-forms) should,
more appropriately, be denoted by $\Pi T^{\ast }.$} $T^{\ast }(\mathbb{R}%
^{(n|m)})$ acting superlinearly on the parity reversed tangent bundle $\Pi T(%
\mathbb{R}^{(n|m)})$, and its integral is defined as follows:
\begin{equation}
I[\omega ]\equiv \int_{\mathbb{R}^{(n|m)}}\omega ^{(n|m)}\equiv
\int_{T^{\ast }(\mathbb{R}^{(n|m)})=\mathbb{R}^{(n+m|m+n)}}\omega (x,\theta
,dx,d\theta )[dxd\theta ~d(dx)d(d\theta )]  \label{Idiomega}
\end{equation}%
where the order of the integration variables is kept fixed. The symbol \-\ $%
[dxd\theta ~d(dx)d(d\theta )]$ denotes the Berezin integration
\textquotedblleft measure" and it is invariant under any coordinate
transformation on $\mathbb{R}^{(n|m)}$. It is a section of the \textit{%
Berezinian bundle} of $T^{\ast }(\mathbb{R}^{(n|m)})$ (a super line bundle
that generalizes the determinant bundle of a purely bosonic manifold). The
sections of the determinant bundle transform with the determinant of the
jacobian and the sections of the Berezinian with the superdeterminant of the
super-Jacobian. The integrations over the fermionic variables $\theta $ and $%
dx$ are Berezin integrals\footnote{%
In the following, for a given set $\{\xi ^{i}\}_{i=1}^{n}$ of Grassmann
variables, our definition of the Berezin integral is $\int \xi ^{1}...\xi
^{n}\left[ d^{n}\xi \right] =1$ and not $\int \xi ^{1}...\xi ^{n}\left[
d^{n}\xi \right] =\left( -1\right) ^{\frac{n(n-1)}{2}}.$ Moreover, if $%
\alpha $ is a monomial expression of some anticommuting variables $\alpha
^{k}$ not depending on the $\xi ^{i},$ we define: $\int \alpha \xi
^{1}...\xi ^{n}\left[ d^{n}\xi \right] =\alpha ,$ where the product between $%
\alpha $ and the $\xi ^{i}$ is the usual $\mathbb{Z}_{2}-$ graded wedge
product in the superalgebra generated by the graded tensor product of the
Grassmann algebra generated by the $\xi ^{i}$ and that generated by the $%
\alpha ^{k}:$ if $\mathcal{A}$ and $\mathcal{B}$ are two $\mathbb{Z}_{2}$%
-graded algebras with products $\cdot _{\mathcal{A}}$and $\cdot _{\mathcal{B}%
}$, the $\mathbb{Z}_{2}$-graded tensor product $\mathcal{A}\otimes \mathcal{B%
}$ is a $\mathbb{Z}_{2}$-graded algebra with the product (for homogeneous
elements) given by :%
\[
(a\otimes b)\cdot _{\mathcal{A}\otimes \mathcal{B}}(a^{\prime }\otimes
b^{\prime })=(-1)^{\left\vert a^{\prime }\right\vert \left\vert b\right\vert
}a\cdot _{\mathcal{A}}a^{\prime }\otimes b\cdot _{\mathcal{B}}b^{\prime }
\]%
In our case the algebras are Grassmann algebras and the products $\cdot $
are wedge products. The symbols $\otimes $ and $\wedge $ will be, in
general, omitted.}, and those over the bosonic variables $x$ and $d\theta $
are Lebesgue integrals (we assume that $\omega (x,\theta ,dx,d\theta )$ has
compact support in the variables $x$ and it is a product of Dirac's delta
distributions in the $d\theta $ variables). A similar approach for a
superform would not be possible because the polynomial dependence on the $%
d\theta $ leads to a divergent integral.

As usual, this definition can be extended to supermanifolds $\mathcal{M}%
^{(n|m)}$ by using bosonic partitions of unity.

Note that this definition of integration is a simple generalization of the
integration of differential forms. For example, if $\omega =f(x)dx$ is an
integrable one form, its integral over $\mathbb{R}$ can be interpreted as a
Berezin integral of the superfunction $\omega (x,dx)=f(x)dx$ on $T^{\ast }(%
\mathbb{R})$ (considered as a supermanifold) with respect to the bosonic
variable $x$ and the fermionic variable $dx:$%
\begin{equation}
I[\omega ]\equiv \int_{\mathbb{R}}\omega \equiv \int_{\mathbb{R}%
}f(x)dx=\int_{T^{\ast }(\mathbb{R}^{(1|0)})=\mathbb{R}^{(1|1)}}\omega
(x,dx)[dxd(dx)]  \label{Idiforma}
\end{equation}%
The symbol $[dxd(dx)]$ denotes the integration "measure" and it is invariant
under any coordinate transformations on $\mathbb{R}$: for a change of
coordinates in $\mathbb{R}$ given by $x=x(y)$ the super-Jacobian matrix is $%
\begin{pmatrix}
\frac{\partial x}{\partial y} & 0\\
0 & \frac{\partial dx}{\partial dy}%
\end{pmatrix}$ whose superdeterminant is $1$ $\left( \text{using }\mathrm{%
sdet}\begin{pmatrix}
A & 0\\
0 & B
\end{pmatrix}=\frac{\mathrm{\det A}}{\mathrm{\det B}}\right) .$

See again Witten \cite{Witten:2012bg} for a more detailed discussion on the
symbol $[dxd\theta d(dx)d(d\theta)]$ and many other important aspects of the
integration theory of integral forms.

According to the previous discussion, if a superform $\omega^{(n|0)}$ with
form degree $n$ (equal to the bosonic dimension of the reduced bosonic
submanifold\footnote{%
see sec.2.} $\mathcal{M}^{(n)}\hookrightarrow \mathcal{M}^{(n|m)})$ and
picture number zero is multiplied by a $(0|m)$ integral form $\gamma^{(0|m)}$%
, we can define the integral on the supermanifold of the product:%
\begin{equation}
\int_{\mathcal{M}^{(n|m)}}\omega^{(n|0)}\wedge\gamma^{(0|m)}.
\label{Iprodotto}
\end{equation}
This type of integrals can be given a geometrical interpretation in terms of
the reduced bosonic submanifold $\mathcal{M}^{(n)}$ of the supermanifold and
the corresponding Poincar\'{e} dual.

We start with a submanifold $\mathcal{S}$ of dimension $s$ of a
differentiable manifold $\mathcal{M}$ of dimension $n.$ We take an embedding
$i:$ 
\[
i:\mathcal{S}\rightarrow \mathcal{M}
\]%
and a compact support $s-$form $\omega \in \Omega ^{s}(\mathcal{M}).$ The
\textbf{Poincar\'{e} dual} of $\mathcal{S}$ is a \textit{closed} form $\eta
_{\mathcal{S}}\in \Omega ^{n-s}(\mathcal{M})$ such that:%
\begin{equation}
I[\omega ,\mathcal{S}]=\int_{\mathcal{S}}i^{\ast }\omega =\int_{\mathcal{M}%
}\omega \wedge \eta _{\mathcal{S}}  \label{catA}
\end{equation}%
where $i^{\ast }$ is the pull-back of forms.

If we suppose that the submanifold $\mathcal{S}$ is described locally by the
vanishing of $n-s$ coordinates $t^{1},\dots,t^{n-s}$, its Poincar\'{e} dual
can also be described as a \textit{singular closed localization form}:%
\begin{equation}
\eta_{_{\mathcal{S}}}=\delta(t^{1})...\delta(t^{n-s})dt^{1}{}\wedge...{}%
\wedge dt^{n-s}   \label{catB}
\end{equation}
This distribution-valued form is clearly closed (from the properties of the
delta distributions $d\,\delta(t)=\delta^{\prime}(t)dt$ and from $%
dt^{i}\wedge dt^{i}=0$). This form belongs to $\Omega^{n-s}(\mathcal{M)}$
and is constructed in such a way that it projects on the submanifold $%
t^{1}=\dots=t^{n-s}=0$. Thus, by multiplying a given form $\omega\in\Omega
^{s}(\mathcal{M)}$ by $\eta_{S}$, the former is restricted to those
components which are not proportional to the differentials $dt^{i}$.

Observing that the Dirac $\delta$-function of an odd variable ($dt$ is odd
if $t$ is even) coincides with the variable itself, we rewrite $\eta _{%
\mathcal{S}}$ in a form that will turn out to be useful for the
generalization to supermanifolds (omitting as usual wedge symbols):
\begin{equation}
\eta_{_{\mathcal{S}}}=\delta(t^{1})...\delta(t^{n-s})\delta(dt^{1}){}...{}%
\delta(dt^{n-s})   \label{catC}
\end{equation}
which heuristically corresponds to the localisation to $t^{1}=\dots=t^{n-s}=0
$ and $dt^{1}=\dots=dt^{n-s}=0$. Note that if a submanifold $\mathcal{S}$ is
described by the vanishing of $n-s$ functions $f^{1}(t)=\dots=f^{n-s}(t)=0$
the corresponding Poincar\'{e} dual $\eta_{_{\mathcal{S}}}$ is:
\begin{equation}
\eta_{_{\mathcal{S}}}=\delta(f^{1})...\delta(f^{n-s})\delta(df^{1}){}...%
\delta(df^{n-s})   \label{catC1}
\end{equation}

If we change (in the same homology class) the submanifold $\mathcal{S}$ to $%
\mathcal{S}^{\prime}$, which is equivalent to change continously the
embedding, the corresponding Poincar\'{e} duals $\eta_{_{\mathcal{S}}}$ and $%
\eta_{_{\mathcal{S}^{^{\prime}}}}$ differ by an \textit{exact} form. This
can be easily proved by recalling that the Poincar\'{e} duals are closed $%
d\eta_{S}=0$ and any variation (denoted by $\Delta$) of $\eta_{\mathcal{S}}$
is exact:
\begin{equation}
\Delta\eta_{_{\mathcal{S}}}=d\Big(\Delta f\delta(f)\Big)   \label{catD}
\end{equation}
Given the explicit expression of $\eta_{_{\mathcal{S}}}$, it is easy to
check eq. (\ref{catD}) by expanding both members assuming that the
derivation $\Delta$ follows the Leibniz rule, and using also the commutation
relation $d\Delta=\Delta d$. For example, in the simple case $\eta_{_{%
\mathcal{S}}}=\delta(f)df$ of a single bosonic function $f$ , we have $\Delta%
\left[ \delta(f)df\right] =\delta^{\prime}(f)\Delta fdf+\delta(f)\Delta df$
, which is also equal to $d\Big(\Delta f\delta(f)\Big)=\Delta
df\delta(f)+\Delta f\delta^{\prime}(f)df.$

Using this property we can show that, if $d\omega=0$ (in $\mathcal{M}$ since
$d_{\mathcal{S}}\left( i^{\ast}\omega\right) =0$ trivially in $\mathcal{S}$%
), then the integral does not depend on the embedding of the submanifold.
Indeed varying the embedding amounts to vary the Poincar\'{e} dual, so that
the variation of the integral reads
\begin{equation}
\Delta I[\omega,\mathcal{S}]=I[\omega,\Delta\mathcal{S}]=\int_{\mathcal{M}%
}\omega\wedge\Delta\eta_{\mathcal{S}}=\int_{\mathcal{M}}\omega\wedge d\xi_{%
\mathcal{S}}=(-)^{s}\int_{\mathcal{M}}d\omega\wedge\xi_{\mathcal{S}}
\label{stokke}
\end{equation}
where $\Delta\eta_{\mathcal{S}}=d\xi_{\mathcal{S}}$.

The same arguments apply in the case of supermanifolds. Consider a
submanifold $\mathcal{S}^{\left( s|q\right) }$ of a supermanifold $\mathcal{M%
}^{\left( n|m\right) }.$ We take an embedding $i:$%
\[
i:\mathcal{S}^{\left( s|q\right) }\rightarrow\mathcal{M}^{\left( n|m\right) }
\]
and an integral form $\omega\in\Omega^{s|q}(\mathcal{M}^{\left( n|m\right) })
$ (integrable in the sense of superintegration when pulled back on $\mathcal{%
S}^{\left( s|q\right) }$). The \textbf{Poincar\'{e} dual} of $\mathcal{S}%
^{\left( s|q\right) }$ is a $d$-\textit{closed} form $\eta_{\mathcal{S}%
}\in\Omega^{n-s|m-q}(\mathcal{M}^{\left( n|m\right) })$ such that:%
\begin{equation}
\int_{\mathcal{S}^{\left( s|q\right) }}i^{\ast}\omega=\int_{\mathcal{M}%
^{\left( n|m\right) }}\omega\wedge\eta_{\mathcal{S}}   \label{eta2}
\end{equation}
Again we can write:%
\begin{equation}
\eta_{_{S}}=\delta(f^{1})...\delta(f^{...})\delta(df^{1})...\delta(df^{...})
\label{eta3}
\end{equation}
where the $f$'s are the functions defining (at least locally) the
submanifold $\mathcal{S}^{\left( s|q\right) }$. Here some of them are even
functions and some of them are odd functions. The Poincar\'{e} dual is a
closed integral form that, in general, if written explicitely in the
coordinates $(x,\theta)$, contains delta forms and their derivatives.%
\footnote{%
We recall that the modules of integral forms are constructed in terms of
compact-support distributions of $d\theta$'s and its derivatives. Therefore
a PCO could in principle contain also the derivatives of Dirac delta forms $%
\delta(d\theta)$. In the forthcoming sections, we will illustrate this point
with explicit examples of PCO's built with derivatives of delta forms. Note
that, instead, the Heaviside (step) function $\Theta(d\theta)$ is not an
admissible distribution for an integral form.}

Again it is easy to check, for example in the simple case of a single $f$
fermionic, that any variation of $\eta_{S}$ is $d$-exact:
\begin{equation}
\Delta\eta_{S}=d\Big((\Delta f)f\delta^{^{\prime}}(df)\Big)   \label{catD-1}
\end{equation}
Note that the two formulae (\ref{catD}) and (\ref{catD-1}) for the variation
of $\eta_{S}$ can be combined in a formula that holds true in both cases:
\begin{equation}
\Delta\eta_{S}=d\Big(\Delta f\delta(f)\delta^{^{\prime}}(df)\Big)
\label{catD-1-1}
\end{equation}
Indeed, one has $\delta^{^{\prime}}(df)=1$ or $\delta(f)=f$ when $f$ is
respectively bosonic or fermionic.

If we take now an embedding $i$ of the reduced bosonic submanifold $\mathcal{%
M}^{(n)}\overset{i}{\hookrightarrow}\mathcal{M}^{(n|m)}$ and a
representative of its Poincar\'{e} dual ${\mathbb{Y}}^{(0|m)}$, we have:
\begin{equation}
\int_{\mathcal{M}^{(n|m)}}\omega^{(n|0)}\wedge{\mathbb{Y}}^{(0|m)}=\int_{%
\mathcal{M}^{(n)}}i^{\ast}\omega^{(n|0)}
\end{equation}

The "standard" embedding is given by $\theta^{\alpha}=0$ for all $\alpha.$
The corresponding standard Poincar\'{e} dual is ${\mathbb{Y}}%
_{st}^{(0|m)}=\theta^{1}...\theta^{m}\delta(d\theta^{1})\dots\delta(d%
\theta^{m}).$

If $\omega^{(n|0)}$ is a \textit{closed} form, this integral is unchanged if
we modify the embedding by adding an exact term to ${\mathbb{Y}}^{(0|m)}$.
This fact will be used to change the standard Poincar\'{e} dual into another
with manifest supersymmetry.

For rigid supersymmetric models, the closed form $\omega^{(n|0)}$ is the
Lagrangian of the model $\mathcal{L}^{(n|0)}(\Phi,V,\psi)$ built using the
rheonomic rules (see \cite{cube}) and it contains the dynamical fields $\Phi$
(each dynamical field is promoted to a superfield) and the rigid
supervielbeins $V^{a}=dx^{a}+\theta\gamma^{a}d\theta\,,\psi^{a}=d\theta^{a}$
satisfying the Maurer-Cartan equations
\begin{equation}
dV^{a}=\psi\gamma^{a}\psi\,,~~~~~d\psi^{a}=0\,.
\end{equation}
In the present formula, we have used real Majorana spinors.

On the other side the Poincar\'{e} dual forms ${\mathbb{Y}}^{(0|m)}$, called
\textit{Picture Changing Operators} (PCO's) in string theory literature (see
\cite{Belo} for details), contains only geometric data (for instance the
supervielbeins or the coordinates themselves).

For rigid supersymmetric models we have
\begin{equation}
S_{rig}=\int_{\mathcal{M}^{(n|m)}}\mathcal{L}^{(n|0)}(\Phi,V,\psi )\wedge{%
\mathbb{Y}}^{(0|m)}(V,\psi)
\end{equation}
with $d\mathcal{L}^{(n|0)}(\Phi,V,\psi)=0$ in order to be able to freely
change the PCO by exact terms, without changing $S_{rig}$.

In the case of supergravity, the supervielbeins $V^{a}$ and $\psi^{\alpha}$
are promoted to dynamical fields $(E^{a},E^{\alpha})$ and therefore the
action becomes
\begin{equation}
S_{sugra}=\int_{\mathcal{M}^{(n|m)}}\mathcal{L}^{(n|0)}(\Phi,E)\wedge {%
\mathbb{Y}}^{(0|m)}(E).
\end{equation}
The closure of the action and of the PCO's implies the conventional
constraints for supergravity, reducing the independent fields to the
physical fields.

\section{PCO's and their Algebraic Properties.}

In this section we recall a few definitions and useful computations about
the PCO's in our notations. For more details see \cite{Belo} and \cite{LMP}.

We start with the \textit{Picture Lowering Operators} that map cohomology
classes in picture $q$ to cohomology classes in picture $r<q.$

Given an integral form, we can obtain a superform by acting on it with
operators decreasing the picture number. Consider the following operator:
\begin{equation}
\delta(\iota_{D})=\int_{-\infty}^{\infty}\exp\Big(it\iota_{D}\Big)dt
\end{equation}
where $D$ is an odd vector field on $T(\mathcal{SM})$ with $\left\{
D,D\right\} \neq0\footnote{%
Here and in the following $\left\{ ,\right\} $ is the anticommutator (i.e.
the graded commutator).}$ and $\iota_D$ is the contraction along the vector $%
D$. The contraction $\iota_D$ is an even operator.

For example, if we decompose $D$ on a basis $D=D^{\alpha}\partial
_{\theta^{\alpha}}$, where the $D^{\alpha}$ are even coefficients and $%
\left\{ \partial_{\theta^{\alpha}}\right\} $ is a basis of the odd vector
fields, and take $\omega=\omega_{\beta}d\theta^{\beta}\in\Omega^{(1|0)}$, we
have
\begin{equation}
\iota_{D}\omega=D^{\alpha}\omega_{\alpha}=D^{\alpha}\frac{\partial\omega }{%
\partial d\theta^{\alpha}}\in\Omega^{(0|0)}\,.
\end{equation}
In addition, due to $\left\{ D,D\right\} \neq0$, we have also that $%
\iota_{D}^{2}\neq0$. The differential operator $\delta(\iota_{\alpha})\equiv%
\delta\left( \iota_{D}\right) $ -- with $D=\partial_{\theta^{\alpha}}$ --
acts on the space of integral forms as follows (we neglect the possible
introduction of derivatives of delta forms, but that generalization can be
easily done):
\begin{align}
\delta(\iota_{\alpha})\prod_{\beta=1}^{m}\delta(d\theta^{\beta}) & =\pm
\int_{-\infty}^{\infty}\exp\Big(it\iota_{\alpha}\Big)\delta(d\theta^{\alpha
})\prod_{\beta=1\neq\alpha}^{m}\delta(d\theta^{\beta})dt  \label{exaG} \\
& =\pm\int_{-\infty}^{\infty}\delta(d\theta^{\alpha}+it)\prod_{\beta
=1\neq\alpha}^{m}\delta(d\theta^{\beta})dt=\mp
i\prod_{\beta=1\neq\alpha}^{m}\delta(d\theta^{\beta})  \nonumber
\end{align}
where the sign $\pm$ is due to the anticommutativity of the delta forms and
it depends on the index $\alpha.$ We have used also the fact that $\exp \Big(%
it\iota_{\alpha}\Big)$ represents a finite translation of $d\theta ^{\alpha}$%
. The result contains $m-1$ delta forms, and therefore it has picture $m-1$.
It follows that $\delta(\iota_{\alpha})$ is an odd operator.

We can define also the Heaviside step operator $\Theta\left( \iota
_{D}\right) $ :%
\begin{equation}
\Theta\left( \iota_{D}\right)
=\lim_{\epsilon\rightarrow0^{+}}-i\int_{-\infty}^{\infty}\frac{1}{t-i\epsilon%
}\exp\Big(it\iota_{D}\Big)dt   \label{ThetasuF}
\end{equation}
The operators $\delta\left( \iota_{D}\right) $ and $\Theta\left( \iota
_{D}\right) $ have the usual formal distributional properties: $\iota
_{D}\delta(\iota_{D})=0$ , $\iota_{D}\delta^{\prime}(\iota_{D})=-\delta
(\iota_{D})$ and $\iota_{D}\Theta\left( \iota_{D}\right) =\delta(\iota _{D}).
$

In order to map cohomology classes into cohomology classes decreasing the
picture number, we introduce the operator (see \cite{Belo}):%
\begin{equation}
Z_{D}=\left[ d,\Theta\left( \iota_{D}\right) \right]   \label{Zeta}
\end{equation}

In the simplest case $D=\partial_{\theta^{\alpha}}$ we have:%
\begin{equation}
Z_{\partial_{\theta^{\alpha}}}=i\delta(\iota_{\alpha})\partial_{\theta
^{\alpha}} \equiv Z_{\alpha}   \label{Zetaalfa}
\end{equation}
The operator $Z_{\alpha}$ is the composition of two operators acting on
different quantities: $\partial_{\theta^{\alpha}}$ acts only on functions,
and $\delta(\iota_{\alpha})$ acts only on delta forms.

In order to further reduce the picture we simply iterate operators of type $Z
$.

In the simple case of $\mathbb{R}^{(1|1)}$ the operator $Z_{1}\equiv Z$ acts
on the spaces $\Omega^{(0|1)}$ and $\Omega^{(1|1)}$ producing elements of $%
\Omega^{(0|0)}$ and $\Omega^{(1|0)}$ respectively.

A generic form $\omega$ $\in\Omega^{(0|1)}\oplus$ $\Omega^{(1|1)}$ but $%
\notin\ker Z$ can be written as:%
\begin{equation}
\omega(x,dx,\theta,d\theta)=f(x)\theta\delta(d\theta)+g(x)\theta
dx\delta(d\theta).
\end{equation}
because $Z\Big(\delta(d\theta)\Big)=Z\Big(dx\delta^{\prime}(d\theta )\Big)=Z%
\Big(\theta dx\delta^{\prime}(d\theta)\Big)=0$

The action of the operator $Z$ is:%
\begin{equation}
Z(\omega)=f(x)-g(x)dx\in\Omega^{(0|0)}\oplus\Omega^{(1|0)}
\end{equation}

As explained in \cite{Witten:2012bg}, the operator $Z$ can be defined also
in terms of "integration along the fibers". Intuitively, to remove a Dirac
delta of a given $d\theta$ from an integral form, changing its picture
number, it is sufficient to integrate along that coordinate.

For example, again in the simple case of $\mathbb{R}^{(1|1)},$ the
transformation of coordinates generated by the vector $\partial_{\theta}$ is
given by:%
\begin{align}
x & \rightarrow x \\
\theta & \rightarrow\theta+\epsilon
\end{align}

where $\epsilon $ is an auxiliary fermionic parameter (see sec. 2 for a
rigorous treatment in terms of the functor of points) .

This change of coordinates maps $\omega$ to
\begin{equation}
\omega^{\ast}=f(x)\left( \theta+\epsilon\right) \delta(d\theta
+d\epsilon)+g(x)\left( \theta+\epsilon\right) dx\delta(d\theta+d\epsilon)
\end{equation}
The picture changing can be obtained integrating with respect to the
variables $d\epsilon$ and $\epsilon:$%
\begin{equation}
Z(\omega)=\int\omega^{\ast}\left[ d\left( d\epsilon\right) d\epsilon \right]
\end{equation}
A similar description in terms of the Voronov integral transform can be
found in \cite{LMP}.

The $Z$ operator is in general not invertible but it is possible to find a
non unique operator $Y$ such that $Z\circ Y$ is an isomorphism in the
cohomology. These operators are the called \textit{Picture Raising
Operators. }The operators of type $Y$ are non trivial elements of the de
Rham cohomology.

We apply a PCO of type $Y$ on a given form by taking the graded wedge
product; given $\omega$ in $\Omega^{(p|q)}$, we have:
\begin{equation}
\omega\overset{Y}{\longrightarrow}\omega\wedge Y\in\Omega^{\left( {p|q+1}%
\right) }\,,   \label{PCOc}
\end{equation}
Notice that if $q=m$, then $\omega\wedge Y=0$. In addition, if $d\omega=0$
then $d(\omega\wedge Y)=0$ (by applying the Leibniz rule), and if $%
\omega\neq dK$ then it follows that also $\omega\wedge Y\neq dU$ where $U$
is a form in $\Omega^{(p-1|q+1)}$. So, given an element of the cohomogy $%
H^{(p|q)}$, the new form $\omega\wedge Y$ is an element of $H^{(p|q+1)}.$

For a simple example in $\mathbb{R}^{(1|1)}$ we can consider the PCO $%
Y=\theta\delta\left( d\theta\right) $, corresponding to the vector $%
\partial_{\theta}$; we have $Z\circ Y=Y\circ Z=1$

More general forms for $Z$ and $Y$ can be constructed, for example starting
with the vector $Q=\partial_{\theta}+\theta\partial_{x}.$

The corresponding PCO of type $Z$ can be computed observing that the
transformation of coordinates generated by the vector $Q=\partial _{\theta
}+\theta \partial _{x}$ is:%
\begin{align}
x& \rightarrow x+\epsilon \theta   \label{sup1} \\
\theta & \rightarrow \theta +\epsilon   \label{sup2}
\end{align}%
If, as usual, we want to consider $\delta _{\epsilon }\theta =\epsilon \,$as
a translation in the (unique) fermionic direction $\theta $ we must conclude
that $\epsilon \theta =0.$ So, if we want to give the geometrical meaning of
a translation to the transformation $\delta _{\epsilon }x=\epsilon \theta $
we should introduce an auxiliary Grassmann algebra with two nilpotents
generators $\epsilon _{1}$ and $\epsilon _{2}.$ In this way $\epsilon \ $and
$\theta $ are both interpreted, using the functor of points, as linear
combinations of $\epsilon _{1}$ and $\epsilon _{2},$ and hence $\epsilon $
and $\theta $ are as usual anticommuting and nilpotent; moreover $\epsilon
\theta $ is not a real number but it is bosonic and different from zero.

This change of coordinates maps the generic integral form $%
\varphi=g(x)\theta dx\delta(d\theta)\in\Omega^{(1|1)}$ to
\begin{equation}
\varphi^{\ast}=g(x+\epsilon\theta)\left( \theta+\epsilon\right) \left(
dx+d\epsilon\theta-\epsilon d\theta\right) \delta(d\theta+d\epsilon)
\end{equation}
The picture changing operation can be obtained integrating over the
variables $d\epsilon$ and $\epsilon:$%
\begin{equation}
Z_{Q}(\varphi)=\int\varphi^{\ast}\left[ d\left( d\epsilon\right) d\epsilon%
\right] =-g(x)dx
\end{equation}
The explicit computation using instead the formula $Z=[d,\Theta(\iota_{Q})]$
is:
\begin{align}\label{ThetasuFiAA}
Z_{Q}[\varphi] & =d[\Theta(\iota_{Q})\varphi]=d\Big[\Theta(\iota
_{Q})g(x)\theta dx\delta(d\theta)\Big] \\
& =d\Big[\lim_{\epsilon\rightarrow0^{+}}-i\int_{-\infty}^{\infty}\frac{1}{%
t-i\epsilon}g(x)\theta dx\delta (d\theta+it)dt\Big]=  \nonumber
 \\
& =d\left[ -\frac{g(x)\theta dx}{d\theta}\right] =-g(x)dx\,.  \nonumber
\end{align}
The last expression is clearly closed. Note that in the above computations 
(\ref{ThetasuFiAA}) we have introduced formally the inverse of the (commuting)
superform $d\theta.$ Using a terminology borrowed from superstring theory we
can say that, even though in a computation we need an object that lives in
the \textit{Large Hilbert Space}, the result is still in the \textit{Small
Hilbert Space}.

Note that the negative powers of the superform $d\theta$ are well defined
only in the complexes of superforms (i.e. in picture $0)$. In this case the
inverse of the $d\theta$ and its powers are closed and exact and behave with
respect to the graded wedge product as \textit{negative degree} superforms
of picture $0$. In picture $\neq0$ negative powers are not defined because
of the distributional relation $d\theta\delta\left( d\theta\right) =0.$

An "inverse" PCO of type $Y$ invariant under the rigid supersymmetry
transformations (generated by the vector $Q$) $\delta_{\epsilon}x=\epsilon%
\theta$ and $\delta_{\epsilon}\theta=\epsilon$ is, for example, given by:%
\begin{equation}
Y_{Q}\mathbb{=}(dx+\theta d\theta)\delta^{\prime}(d\theta)
\end{equation}
We have:
\begin{equation}
Y_{Q}Z_{Q}[\varphi]=-g(x)dx\wedge(dx+\theta
d\theta){}\delta^{\prime}(d\theta)=g(x)\theta dx\delta(d\theta)=\varphi\,.
\end{equation}

\section{Super-Quantum Mechanics.}

\subsection{$D=1,N=1.$}

In the present section, we present a very special model in the lowest
possible dimension $D=1$ and $N=1$, namely $N=1$ super quantum mechanics%
\footnote{%
See sec.2.1 for a mathematical introduction to the ``supersymmentric
point particle" and the related formalism and concepts.}. This model is very
useful to understand several details in more complicate theories and
provides a simple and calculable example. We list all ingredients and we
discuss some implications.

\begin{enumerate}
\item The local coordinates of the superspace $\mathbb{R}^{(1|1)}$ are
denoted by $(x,\theta)$, the flat supervielbeins are given by $V=dx+\theta
d\theta$, $\psi=d\theta$, satisfying the usual Maurer-Cartan algebra
\begin{equation}
dV=\psi^{2}\,,~~~~~~d\psi=0\,.
\end{equation}
The covariant derivative $D$ and the supersymmetry generator $Q$ are
\begin{equation}
D=\partial_{\theta}-\theta\partial_{x}\,,~~~~~~~~Q=\partial_{\theta}+\theta%
\partial_{x}\,,~~~~~~~~
\end{equation}
with the algebra
\begin{equation}
D^{2}=\frac{1}{2}\{D,D\}=-\partial_{x}\,,~~~~~~Q^{2}=\frac{1}{2}%
\{Q,Q\}=\partial_{x}\,,~~~~~~\{Q,D\}=0\,.   \label{algebra}
\end{equation}

\item To construct a Lagrangian we need superfields. The supermultiplet in
this simplified framework is composed by a single boson and a single
fermion. Then, we can easily arrange them into a single scalar superfield $%
\Phi$. If we denote by $(\phi,\lambda)$ its component fields, we have
\begin{align}
\Phi(x,\theta) & =\phi(x)+\theta\lambda(x)\,,  \label{dunoC} \\
W(x,\theta) & =D\Phi=\lambda(x)-\theta\partial_{x}\phi(x)\,, \\
F(x,\theta) &
=D\,W(x,\theta)=D^{2}\Phi=-\partial_{x}\Phi=-\partial_{x}\phi(x)-\theta%
\partial_{x}\lambda(x)\,.
\end{align}
Together with the superfield $\Phi$ we have also some derived superfields
such as $W$ and $F$. $W$ plays the r\^{o}le of a fermionic superfield (the
superfield whose first component is the physical fermion $\lambda$).

\item Supersymmetry. If we denote by $\epsilon$ the constant anticommuting
parameter of supersymmetry, we have
\begin{equation}
\delta_{\epsilon}\Phi=\epsilon
Q\Phi=\epsilon(\lambda+\theta\partial_{x}\phi)\,,
\end{equation}
from which we deduce the supersymmetry variations of the component fields
\begin{equation}
\delta_{\epsilon}\phi=\epsilon\lambda\,,~~~~~~~\delta_{\epsilon}\lambda=-%
\partial_{x}\phi\epsilon\,.
\end{equation}
These transformations will be used to check the invariance of the action and
the covariance of the equations of motion.

\item Supersymmetric action and the equations of motion. We write the action%
\footnote{%
The action $S=\frac{1}{2}\int\partial_{x}\Phi D\Phi\lbrack dx\,d\theta]$ is
manifestly invariant under the supersymmetry transformation \ref{sup1} and %
\ref{sup2}.} in the superspace version and then we compute the component
action explicitly (the integrals are usual Berezin integrals):
\begin{align}
S & =\frac{1}{2}\int\partial_{x}\Phi D\Phi\lbrack dx\,d\theta]=\frac{1}{2}%
\int\Big((\partial_{x}\phi+\theta\partial_{x}\lambda)(\lambda
-\theta\partial_{x}\phi)\Big)[dx\,d\theta]  \label{dunoF} \\
& =\frac{1}{2}\int\Big(\partial_{x}\phi\lambda+\theta(\partial_{x}\lambda%
\lambda-(\partial_{x}\phi)^{2})\Big)[dx\,d\theta]=-\frac{1}{2}\int_{\mathbb{R%
}}\Big[\lambda\partial_{x}\lambda+(\partial_{x}\phi )^{2}\Big]dx\,.
\nonumber
\end{align}
and the equations of motion are:
\begin{equation}
\partial_{x}D\Phi=0\,,~~~~~\Longrightarrow~~~~~\partial_{x}^{2}\phi
=0\,,~~~~\partial_{x}\lambda=0\,,   \label{eqcampo}
\end{equation}
i.e. the Klein-Gordon and the Dirac equation. The model is very simple and
there is a complete match of degrees of freedom both off-shell and on-shell.
Thus we do not need auxiliary fields.

\item Let us move to the geometrical construction. For that, we extend all
physical fields to superfields. In our case the field $\phi$ is promoted to $%
\Phi$ and the fermion $\lambda$ is promoted to $W$ (we adopt the same
letters as above because it will turn out that they do coincide). Promoting
the fields to superfields introduces more degrees of freedom, and we reduce
them by imposing the differential conditions
\begin{equation}
d\Phi=V\partial_{x}\Phi+\psi
W\,,~~~~~~dW=V\partial_{x}W-\psi\partial_{x}\Phi\,,   \label{condiff}
\end{equation}
Being $d$ nilpotent, the equation for $W$ is obtained by imposing the
Bianchi identities. From these equations we immediately see that
\begin{equation}
D\Phi=W\,,~~~~~DW=D^{2}\Phi=-\partial_{x}\Phi\,.   \label{antA}
\end{equation}

\item Now, we are in a position to construct the geometric Lagrangian $%
\mathcal{L}^{(1|0)}$. It is built in terms of superforms, their
differentials and geometrical data of the supermanifold parametrized by $%
V,\psi$. Using the Hodge dual in supermanifolds one could construct a
supersymmentric action in the supermanifold using a Lagrangian in picture $1$
of type $\mathcal{L}^{(1|1)}$ (see section 3.2.2 of the paper \cite%
{Castellani:2015ata}). For constructing instead a geometrical Lagrangian in
picture $0$, we need an additional $(0|0)$-form superfield $\xi$ (first
order formalism) and we have
\begin{equation}
\mathcal{L}^{(1|0)}=\frac{1}{2}\xi^{2}V+\xi(d\Phi-\psi W)+\frac{1}{2}WdW\,,
\label{antB}
\end{equation}
from which we compute the equations of motion
\begin{equation}
\xi V+d\Phi-\psi W=0\,,~~~~~~~~d\xi=0\,,~~~~~~-\xi\psi+dW=0\,.   \label{antC}
\end{equation}
The rheonomic action is built according to the rules presented in \cite{cube}%
: inspired by the kinetic terms of the component action, promoting all
fields to superfields and adding those terms allowed by scaling dimensions,
form degree and other quantum numbers. Then, imposing the $d$-closure one
fixes the coefficients. The equations stemming from that action should
reproduce both the differential conditions (\ref{condiff}) and the equations
of motion (\ref{eqcampo}).

By expanding $d\Phi=V\partial_{x}\Phi+\psi D\Phi$ and $dW=V\partial_{x}W+%
\psi D\partial_{x}W$, we have
\begin{equation}
\xi=-\partial_{x}\Phi\,,~~~~~D\Phi=W\,,~~~~~\partial_{x}\xi=0\,,~~~~~%
\partial _{x}W=0\,,~~~~~DW=\xi\,.   \label{antD}
\end{equation}
so that by consistency
\[
\partial_{x}^{2}\Phi=0\,,~~~~~DW=-\partial_{x}\Phi\,.
\]
These equations are the complete set of conditions and equations of motion.
For the convenience of reader, we also decompose the action into the $V$%
-dependent part and the $\psi$-dependent part
\begin{equation}
\mathcal{L}^{(1|0)}=\Big(\frac{1}{2}\xi^{2}+\xi\partial_{x}\Phi-\frac{1}{2}%
W\partial_{x}W\Big)V+\Big(\xi(D\Phi-W)+\frac{1}{2}WDW\Big)\psi\,.
\label{antE}
\end{equation}

\item Closure of the action. As discussed in sec.3, in order to be able to
choose freely the appropriate PCO, the action must be closed under $d$.
Notice that in a supermanifold this statement is not trivial. We have
\begin{align}
d\mathcal{L}^{(1|0)}& =d\xi \wedge (\xi V+d\Phi -\psi W)+\frac{1}{2}\xi
^{2}\psi \wedge \psi -\xi dW\wedge \psi +\frac{1}{2}dW\wedge dW
\label{dunoN} \\
& =d\xi \wedge (\xi V+d\Phi -\psi W)+\frac{1}{2}\Big(\xi \psi -dW\Big)\wedge %
\Big(\xi \psi -dW\Big).  \nonumber
\end{align}%
Using the first equation of motion, namely $\xi V+d\Phi -\psi W=0$, we get
that the first term vanishes. To prove the vanishing of the second term, one
needs the last equation in (\ref{antC}). Since these equations of motion are
algebraic, we can use them at the level of the action. This proves that the
action is closed without using the auxiliary fields.

\item PCO's. As described in sec. 3, we construct the action integral for
the supermanifold by the formula
\begin{equation}
S_{geo}=\int_{\mathbb{R}^{(1|1)}}\mathcal{L}^{(1|0)}\wedge {\mathbb{Y}}^{{%
(0|1)}}\,,
\end{equation}%
where ${\mathbb{Y}}^{{(0|1)}}$ is a PCO. We consider here two possible
choices (where $st$ means standard and $ss$ means supersymmetric)
\begin{equation}
{\mathbb{Y}}_{st}^{(0|1)}=\theta \delta (\psi )\,,~~~~~~~{\mathbb{Y}}%
_{ss}^{(0|1)}=-V\delta ^{\prime }(\psi )\,,
\end{equation}%
which are both closed and not exact. The first one is not manifestly
supersymmetric, but its variation under a supersymmetry transformation is $d$%
-exact. The second one is supersymmetric invariant. The two PCO's are
cohomologous:%
\begin{equation}
{\mathbb{Y}}_{st}^{(0|1)}-{\mathbb{Y}}_{ss}^{(0|1)}=d\left( x\delta ^{\prime
}(\psi )\right)
\end{equation}

\item Component action. Choosing ${\mathbb{Y}}_{st}^{(0|1)}$ we have:
\begin{align}
S & =\int_{\mathbb{R}^{(1|1)}}\mathcal{L}^{(1|0)}\wedge{\mathbb{Y}}%
_{st}^{(0|1)}  \label{dunoP} \\
& =\int_{T^{\ast}\mathbb{R}^{(1|1)}}\Big(\frac{1}{2}\xi^{2}+\xi\partial
_{x}\Phi-\frac{1}{2}W\partial_{x}W\Big)V\theta\delta(\psi)[dxd\theta
d(dx)d\psi]  \nonumber \\
& =\int_{T^{\ast}\mathbb{R}^{(1|1)}}\Big(\frac{1}{2}\xi_{0}^{2}+\xi
_{0}\partial_{x}\phi-\frac{1}{2}\lambda\partial_{x}\lambda\Big)dx\theta
\delta(\psi)[dxd\theta d(dx)d\psi]  \nonumber
\end{align}
where $\xi_{0}=\xi(x,0)$ (namely the first component of superfield $\xi$).
Now, we can integrate over $\theta,$ $dx$ and $\psi$ to get the final
component action (\ref{dunoF}).

\item Superspace action. Choosing ${\mathbb{Y}}_{ss}^{(0|1)}$ we have to
pick up the second term in (\ref{antE}), because of the derivative of the
Dirac delta function. Then, we have
\begin{align}
S & =\int_{\mathbb{R}^{(1|1)}}\mathcal{L}^{(1|0)}\wedge{\mathbb{Y}}%
_{ss}^{(0|1)}  \label{dunoQ} \\
& =-\int_{T^{\ast}\mathbb{R}^{(1|1)}}\Big(\xi(D\Phi-W)+\frac{1}{2}WDW\Big)%
\psi V\delta^{\prime}(\psi)[dxd\theta d(dx)d\psi]  \nonumber \\
& =\int_{T^{\ast}\mathbb{R}^{(1|1)}}\Big(\xi(D\Phi-W)+\frac{1}{2}WDW\Big)%
V\delta(\psi)[dxd\theta d(dx)d\psi]\,.  \nonumber
\end{align}
The equation of motion for $\xi$ implies that $W=D\Phi$ and, being an
algebraic equation, we can insert it back into the action and obtain
\begin{equation}
S=\frac{1}{2}\int_{\mathbb{R}^{(1|1)}}\partial_{x}\Phi D\Phi\lbrack
dxd\theta]\,.
\end{equation}
after Berezin integration over the variables $dx$ and $\psi$.

We thus retrieve the superspace action.

\item Picture Lowering Operator. We use the PCO $Z=[d,\Theta(\iota_{\psi})]$
where $\Theta(\iota_{\psi})$ is the Heavisde function (step function) of the
contraction operator $\iota_{\psi}$. Notice that $[d,\iota_{\psi }]f=%
\mathcal{L}_{\psi}f=Df$. Then, we can write $Z$ as $Z=D\delta(\iota_{\psi })$%
. Let us check its action on ${\mathbb{Y}}_{st}^{(0|1)},{\mathbb{Y}}%
_{ss}^{(0|1)}$ and on the volume form $\mathrm{Vol}^{(1|1)}=dx\delta(\psi)$:
\begin{align}
& Z\left( {\mathbb{Y}}_{st}^{(0|1)}\right) =[d,\Theta(\iota_{\psi })]{%
\mathbb{Y}}_{st}^{(0|1)}=d\left( \Theta(\iota_{\psi})\theta\delta
(\psi)\right) =d\left( \frac{\theta}{\psi}\right) =1  \label{dunoR} \\
& Z\left( {\mathbb{Y}}_{ss}^{(0|1)}\right) =[d,\Theta(\iota_{\psi })]{%
\mathbb{Y}}_{ss}^{(0|1)}=-d\left( \Theta(\iota_{\psi})V\delta^{\prime
}(\psi)\right) =-d\left( \frac{V}{\psi^{2}}\right) =-1 \\
& Z\left( \mathrm{Vol}^{(1|1)}\right) =[d,\Theta(\iota_{\psi})]V\delta
(\psi)]=d\left( \Theta(\iota_{\psi})V\delta(\psi)\right) =d\left( \frac {V}{%
\psi}\right) =\psi
\end{align}
Notice that since the two PCO differ by an exact term, that difference drops
out acting with $Z$. Notice also that, again, even though we need an object
living in the \textit{Large Hilbert Space}, the result of the computation
still is in the \textit{Small Hilbert Space}. In addition, using the
definition of the super Hodge dual $\star$ proposed in the paper \cite%
{Castellani:2015ata}, it can be easily seen that $\star\psi={\mathbb{Y}}%
_{ss}^{(1|0)}$, or, equivalently:
\begin{equation}
\psi\wedge-V\delta^{\prime}(\psi)=\mathrm{Vol}^{(1|1)}\,.
\end{equation}
\end{enumerate}

\subsection{$D=1,N=1$ Curved.}

We consider a curved $N=1$ supermanifold $\mathcal{M}^{(1|1)}$ locally
parametrized by the coordinates $(x,\theta).$ Its geometry is described by
the supervielbein
\begin{equation}
E^{v}=E_{x}^{v}V+E_{\theta}^{v}\psi\,,~~~~~~~E^{\psi}=E_{x}^{\psi}V+E_{%
\theta }^{\psi}\psi\,   \label{CIPA}
\end{equation}
where $V$ and $\psi$ are the flat superspace supervielbeins. We impose the
supergravity constraints
\begin{equation}
dE^{v}=E^{\psi}\wedge E^{\psi}\,,~~~~~~dE^{\psi}=0\,.   \label{CIPB}
\end{equation}
which are solved by
\begin{equation}
E^{v}=(E_{\theta}^{\psi})^{2}V\,,~~~~E^{\psi}=E_{\theta}^{\psi}\psi
+DE_{\theta}^{\psi}V\,.   \label{CIPC}
\end{equation}
The geometrical properties of the supermanifold are encoded into the
superfield $E_{\theta}^{\psi}(x,\theta)$. Since a curved supermanifold $%
\mathcal{M}^{(1|1)}$ is rather simple, the torsion constraints (\ref{CIPB})
concentrate the entire geometrical information into a single superfield \cite%
{Sorokin:1999jx}. This will happen also in the $\mathcal{M}^{(1|2)}$ case as
will be shown later.

Because of the equations (\ref{CIPB}), it is easy to show that
\begin{equation}
{\mathbb{Y}}^{(0|1)}\equiv-E^{v}\delta^{\prime}(E^{\psi})
\end{equation}
is a closed integral form. Using the solution (\ref{CIPC}), we have
\begin{equation}
{\mathbb{Y}}^{(0|1)}=-(E_{\theta}^{\psi})^{2}V\delta^{\prime}(\psi )\Big(%
E_{\theta}^{\psi}\psi+DE_{\theta}^{\psi}V\Big)=-(E_{\theta}^{\psi})^{2}V%
\frac{1}{(E_{\theta}^{\psi})^{2}}\delta^{\prime}(\psi)=-V\delta^{\prime
}(\psi)
\end{equation}
which coincides with the flat supersymmetric PCO ${\mathbb{Y}}%
_{ss}^{(0|1)}=-V\delta^{\prime}(\psi)$.

The volume form is in this case
\begin{align}
\mathrm{Vol}^{(1|1)} &=E^{v}\delta(E^{\psi})=(E_{\theta}^{\psi})^{2}V\delta%
\Big(E_{\theta}^{\psi}\psi+DE_{\theta}^{\psi}V\Big)  \nonumber \\
&=E_{\theta}^{\psi }\,V\delta(\psi)=E_{\theta}^{\psi}\,\mathrm{Vol}%
_{flat}^{(1|1)}=\mathrm{Sdet}(E)\mathrm{Vol}_{flat}^{(1|1)}   \label{CIF}
\end{align}
where $\mathrm{Vol}_{flat}^{(1|1)}=V\delta(\psi)$ is the flat volume form.
Note that:
\begin{equation}
E^{\psi}\wedge{\mathbb{Y}}^{(0|1)}=-E^{\psi}\wedge E^{v}\delta^{\prime
}(E^{\psi})=\mathrm{Vol}^{(1|1)}\,.   \label{CIG}
\end{equation}
As a check, we have also $\Big(E_{\theta}^{\psi}\psi+DE_{\theta}^{\psi }V%
\Big)\wedge-V\delta^{\prime}(\psi)=E_{\theta}^{\psi}\mathrm{Vol}%
_{flat}^{(1|1)}=\mathrm{Vol}^{(1|1)}$.

A PCO for a generic (non constrained) vielbein is given by:
\begin{align}
{\mathbb{Y}}^{(0|1)} & =\Big(E_{x}^{v}V+E_{\theta}^{v}\psi\Big)\delta
^{\prime}(E_{x}^{\psi}V+E_{\theta}^{\psi}\psi)  \nonumber \\
&=\Big(E_{x}^{v}V+E_{\theta}^{v}(E_{\theta}^{\psi})^{-1}(\psi-E_{x}^{\psi}V)%
\Big)\frac{1}{(E_{\theta }^{\psi})^{2}}\delta^{\prime}(\psi)  \label{curH} \\
& =\frac{1}{(E_{\theta}^{\psi})^{2}}\Big(E_{x}^{v}-E_{\theta}^{v}(E_{\theta
}^{\psi})^{-1}E_{x}^{\psi}\Big)V\delta^{\prime}(\psi)-E_{\theta}^{v}\frac {1%
}{(E_{\theta}^{\psi})^{3}}\delta(\psi)  \nonumber
\end{align}
If we set $\Big(E_{x}^{v}-E_{\theta}^{v}(E_{\theta}^{\psi})^{-1}E_{x}^{\psi }%
\Big)=0$, $E_{\theta}^{v}=\theta$ and $E_{\theta}^{\psi}=1$, we get the PCO $%
\theta\delta(\psi)$. If we set $\Big(E_{x}^{v}-E_{\theta}^{v}(E_{\theta
}^{\psi})^{-1}E_{x}^{\psi}\Big)=-1$ and $E_{\theta}^{v}=0$, we get instead
the supersymmetric PCO $-V\delta^{\prime}(\psi)$.

\section{Super Quantum Mechanics $N=2$.}

Here we formulate the SQM $N=2$ in the language of integral forms. We follow
the same strategy as in the previous section. We first discuss the
superfield for the multiplet (in the present case, we need also the
auxiliary field $F$ to close the algebra). Then we describe the action in
superspace and the equations of motion. We also give the action in
components and then we study the rheonomic (geometric) action.

To describe the $N=2$ model we recall that we have a scalar field $\phi$,
two fermions $\lambda$ and $\bar{\lambda}$ and an auxiliary field $f$. Both
on-shell and off-shell we get a matching of the fermionic and bosonic
degrees of freedom. The superspace is described by a bosonic coordinate $x$
and two fermionic coordinates $\theta$ and $\bar{\theta}$. The
supersymmetric vielbeins are $V=dx+i(\theta d\bar{\theta}+d\theta\bar{\theta}%
)$, $\psi=d\theta$ and $\bar{\psi}=d\bar{\theta}$ and they satisfy the MC
equations
\begin{equation}
dV=2i\psi\wedge\bar{\psi}\,,~~~~~~~~d\psi=0\,,~~~~~d\bar{\psi}=0\,.
\label{SQMA}
\end{equation}
Correspondently, the superderivatives are defined as
\begin{equation}
D=\partial_{\theta}+i\bar{\theta}\partial_{x}\,,~~~~~\bar{D}=\partial _{\bar{%
\theta}}+i\theta\partial_{x}\,,   \label{SQMAA}
\end{equation}
with the algebra
\begin{equation}
\{D,D\}=0\,,~~~~\{\bar{D},\bar{D}\}=0\,,~~~~~\{D,\bar{D}\}=2i\partial_{x}\,.
\label{SQMAB}
\end{equation}
The physical degrees of freedom are collectively encoded into a $N=2$
superfield $\Phi$ and its derivatives:
\begin{align}
\Phi(x,\theta,\bar{\theta}) & =\phi(x)+\lambda(x)\bar{\theta}+\bar{\lambda }%
(x)\theta+f(x)\theta\bar{\theta}\,,  \label{SQMB} \\
\overline{W}(x,\theta,\bar{\theta}) & =D\Phi=-\lambda+\theta(i\partial
_{x}\phi-f)-i\theta\bar{\theta}\partial_{x}\lambda\,, \\
W(x,\theta,\bar{\theta}) & =-\bar{D}\Phi=\bar{\lambda}+\bar{\theta }%
(i\partial_{x}\phi+f)-i\theta\bar{\theta}\partial_{x}\bar{\lambda}\,.
\end{align}
From these equations, we can compute the product of $D\Phi\bar{D}\Phi$ and
collecting the term proportional to $\theta\bar{\theta}$, we get
\begin{equation}
D\Phi\bar{D}\Phi=\dots+\theta\bar{\theta}\Big[(\partial_{x}\phi)^{2}+i(\bar{%
\lambda}\partial_{x}\lambda-\partial_{x}\bar{\lambda}\lambda )+f^{2}\Big]
\label{SQMC}
\end{equation}
which implies that the superspace action and the component action are given
by
\begin{equation}
S_{N=2,kin}=\frac{1}{2}\int D\Phi\bar{D}\Phi\lbrack dxd\theta d\bar{\theta }%
]=\int\Big[\frac{1}{2}(\partial_{x}\phi)^{2}+i\bar{\lambda}\partial
_{x}\lambda+\frac{f^{2}}{2}\Big]dx   \label{SQMD}
\end{equation}
To include the interaction terms, we consider the function $\mathcal{W}(\Phi)
$ and we add the action
\begin{align}
S_{N=2,int} & =\int\mathcal{W}(\Phi)[dxd\theta d\bar{\theta}]  \label{SQME}
\\
& =\int\bar{D}\Big(\mathcal{W}^{\prime}(\Phi)D\Phi\Big)dx=\int \Big(\mathcal{%
W}^{\prime\prime}\bar{D}\Phi D\Phi+\mathcal{W}^{\prime}\bar {D}D\Phi\Big)dx
\nonumber \\
& =\int\Big(2\mathcal{W}^{\prime}(\phi)f-\mathcal{W}^{\prime\prime}\lambda%
\bar{\lambda}\Big)dx\,.  \nonumber
\end{align}

Now, we consider the geometric approach. For that we start from the
differential of $\Phi$:
\begin{equation}
d\Phi=V\partial_{x}\Phi+\psi\overline{W}-\bar{\psi}W\,,   \label{SQMF}
\end{equation}
where $W=-\bar{D}\Phi$ and $\overline{W}=D\Phi$. Then, we can apply the
differental $d$ on both sides of (\ref{SQMF}) to derive the equations for $W$
and $\bar{W}$. We get
\begin{align}
d\overline{W} & =V\partial_{x}\overline{W}-i\bar{\psi}\partial_{x}\Phi +\bar{%
\psi}F\,, \\
dW & =V\partial_{x}W+i\psi\partial_{x}\Phi+\psi F\,,   \label{SQMG}
\end{align}
where we have introduced the auxiliary superfield $F$ to solve the
consistency condition for the first Bianchi identity. Then, applying again
the differential, we derive the condition on $F:$
\begin{equation}
dF=V\partial_{x}F-i(\psi\partial_{x}\overline{W}+\bar{\psi}\partial_{x}W)\,.
\label{SQMH}
\end{equation}
No additional superfield is needed to close the algebra. From these
relations, we can easily derive the equations relating the four superfields $%
\Phi,W,\overline{W},F$:
\begin{align}
& W=-\bar{D}\Phi\,,~~~~~\bar{W}=D\Phi\,,~~~~~  \label{SQMK} \\
& D\bar{W}=0\,,~~~~~\bar{D}W=0\,, \\
& \overline{D}\overline{W}=F-i\partial_{x}\Phi\,,~~~~DW=F+i\partial_{x}\Phi%
\,.
\end{align}
Again it is easy, by using the algebra of superderivatives, to check that
all the above equations are consistent. We can now construct the geometric
Lagrangian $\mathcal{L}_{N=2}=\mathcal{L}_{N=2,kin}+\mathcal{L}_{N=2,int}$
as follows
\begin{align}
\mathcal{L}_{N=2,kin} & =\xi\Big(d\Phi-\psi\overline{W}+\bar{\psi }W\Big)-%
\frac{1}{2}(\xi^{2}+F^{2})V+\frac{i}{2}\Big(\overline{W}dW+d\overline{W}W%
\Big)\,,  \label{SQMI} \\
\mathcal{L}_{N=2,int} & =\Big(\mathcal{W}^{\prime}F-\mathcal{W}%
^{\prime\prime}W\overline{W}\Big)V-i\mathcal{W}^{\prime}\Big(\psi\overline {W%
}+W\bar{\psi}\Big)
\end{align}
$\mathcal{L}_{N=2}$ is a closed $(1|0)$ form (see equation \ref{SQMM}
below). Notice that, again, the Hodge dual operator has not been used and an
additional superfield $\xi$ is needed in order to write the action in first
order formalism. Note also that there are only three quantities that carry
the $1$-form degree, namely $V,\psi$ and $\bar{\psi}$, and that the action
can be expanded into powers of them. In the present case it is rather easy
since the action is linear in $V,\psi$ and $\bar{\psi}$.

From the Lagrangian we can easily compute the equations of motion which read
(by setting the superpotential $\mathcal{W}=0$ to simplify the discussion)
\begin{align}
& d\Phi-\psi\overline{W}+\bar{\psi}W-\xi V=0\,,~~~~  \label{SQML} \\
& d\xi=0\,,~~~~ \\
& F=0\,, \\
& d\overline{W}=-i\bar{\psi}\xi\,,~~~~ \\
& dW=i\xi\psi\,.
\end{align}
It is an easy exercise to check the consistency of this set of equations.
Then, by computing the differential of $\mathcal{L}_{N=2,kin}$ we get:
\begin{align}
d\mathcal{L}_{N=2,kin} & =d\xi\wedge(d\Phi-\psi\overline{W}+\bar{\psi}W-\xi
V)-FdF\wedge V  \label{SQMM} \\
& +i(d\overline{W}-i\bar{\psi}\xi)\wedge(dW+i\xi\bar{\psi})=0  \nonumber
\end{align}
The differential vanishes because of the algebraic equations of motion (\ref%
{SQML}). In the same way one can show that also the interaction term $%
\mathcal{L}_{N=2,int}$ is closed. The kinetic terms and the interaction
terms are independent, hence the closure property must be shown by taking
the differential of $\mathcal{L}_{N=2,kin}$ and of $\mathcal{L}_{N=2,int}$
separately.\footnote{%
One has to use the equations of motion with the interaction terms.} This is
consistent with what is observed in the component formalism: in the case $N=2
$, one can freely add a superpotential to the action and its parameters are
independent coupling constants. At the level of the rheonomic action we can
add a combination of superfields with the correct scaling dimensions, form
degree and other quantum numbers, which is independent of the kinetic terms.

Before constructing the PCO, we expand the action in powers of the
gravitinos $\psi$ and $\bar{\psi}$. We find:
\begin{align}
\mathcal{L}_{N=2,kin} & =\Big(\xi\partial_{x}\Phi--\frac{1}{2}%
(\xi^{2}+F^{2})+\frac{i}{2}(\overline{W}\partial_{x}W-W\partial_{x}%
\overline {W})\Big)V \\
& +\Big(\xi(D\Phi-\overline{W})-\frac{i}{2}(D\overline{W}W-\overline {W}DW)%
\Big)\psi  \nonumber \\
& +\Big(\xi(\bar{D}\Phi+W)-\frac{i}{2}(\bar{D}\overline{W}W-\overline{W}\bar{%
D}W)\Big)\bar{\psi}\,.  \nonumber \\
&  \nonumber \\
\mathcal{L}_{N=2,int} & =(\mathcal{W}^{\prime}F-\mathcal{W}^{\prime\prime }W%
\overline{W})V-i(\mathcal{W}^{\prime}\overline{W})\psi-i(\mathcal{W}%
^{\prime}W)\bar{\psi}\,,
\end{align}
Each piece of this expansion encodes all information regarding the equations
of motion of the theory. Therefore, by choosing a suitable PCO, one can
derive various equivalent forms of the action with different amounts of
manifest supersymmetries. This would be interesting for applications where
only some partial supersymmetries can be manifestly realized (such as in $D=4
$ $N=4$ SYM).

Now, we are in a position to construct the PCO's. The PCO producing the
component action is the simplest (standard) choice:
\begin{equation}
{\mathbb{Y}}_{st}^{(0|2)}=\theta\bar{\theta}\delta(\psi)\delta(\bar{\psi}%
)\,.   \label{SQMP}
\end{equation}
Then the component action is obtained as
\begin{align}
S_{N=2} & =\int_{\mathcal{M}^{(1|2)}}\mathcal{L}_{N=2}^{(1|0)}\wedge{\mathbb{%
Y}}_{st}^{(0|2)}  \label{SQMQ} \\
& =\int\Big(\xi\partial_{x}\Phi--\frac{1}{2}(\xi^{2}+F^{2})+\frac{i}{2}(%
\overline{W}\partial_{x}W-W\partial_{x}\overline{W})\Big)\theta\bar {\theta}%
[dxd\theta d\bar{\theta}]  \nonumber \\
& +\int\Big(\mathcal{W}^{\prime}F-\mathcal{W}^{\prime\prime}W\overline {W}%
\Big)\theta\bar{\theta}[dxd\theta d\bar{\theta}]\,.  \nonumber
\end{align}
The presence of the $\theta\bar{\theta}$ factor projects all superfields to
their first components and then to the component action.

To reproduce the superspace action, we need another PCO. For that we see
that the following expression
\begin{equation}
{\mathbb{Y}}^{(0|2)}=-\frac{1}{2}iV\wedge(\theta\iota-\bar{\theta}\bar{\iota
})\delta(\psi)\delta(\bar{\psi})\,,   \label{SQMR}
\end{equation}
has the correct properties. The symbols $\iota$ and $\bar{\iota}$ denote the
derivative with respect to $\psi$ and $\bar{\psi}$. Let us compute its
differential
\begin{equation}
d{\mathbb{Y}}^{(0|2)}=i\psi\bar{\psi}\wedge(\theta\iota-\bar{\theta}\bar {%
\iota})\delta(\psi)\delta(\bar{\psi})-iV\wedge(\psi\iota-\bar{\psi}\bar {%
\iota})\delta(\psi)\delta(\bar{\psi})=0\,.   \label{SQMS}
\end{equation}
The first term vanishes because one of the two gravitinos ($\psi$ and $\bar{%
\psi}$) goes through the derivatives $\iota$ and $\bar{\iota}$ until it hits
the corresponding Dirac delta. On the other side the two terms in the second
piece are not vanishing separately: we have to perform an integration by
parts for $\iota$ and $\bar{\iota}$ yielding two identical terms which
cancel each other. To check that ${\mathbb{Y}}^{(0|2)}$ is not exact, we use
the formula
\begin{equation}
d\frac{1}{2}\Big[iV\theta\bar{\theta}\iota\bar{\iota}\delta(\psi)\delta (%
\bar{\psi})\Big]=-i\frac{1}{2}V\wedge(\theta\iota-\bar{\theta}\bar{\iota }%
)\delta(\psi)\delta(\bar{\psi})-\theta\bar{\theta}\delta(\psi)\delta (\bar{%
\psi})=\ {\mathbb{Y}}^{(0|2)}-{\mathbb{Y}}_{st}^{(0|2)}   \label{SQMT}
\end{equation}
which shows that the two PCO's in (\ref{SQMP}) and (\ref{SQMR}) are
cohomologous. The presence of $\theta$ and $\bar{\theta}$ in the expression
is crucial to get the correct superspace action:
\begin{equation}
S_{N=2}=\int_{\mathcal{M}^{(1|2)}}\mathcal{L}_{N=2}^{(1|0)}\wedge{\mathbb{Y}}%
^{(0|2)}\,.   \label{SQMU}
\end{equation}
The contribution to the superspace action comes from the two terms
proportional to the gravitinos $\psi$ and $\bar{\psi}$. The structure of the
PCO (\ref{SQMR}) resembles that in higher dimensions\footnote{%
The applications of the formalism of integral forms to theories in higher
dimensions will be the subject of a forthcoming paper. The case $D=3$ $N=1$
supergravity was analyzed in \cite{3dsuper}.}. The presence of the
superspace coordinates $\theta$ and $\bar{\theta}$ prevents it from being
manifestly supersymmetric. However, as for ${\mathbb{Y}}_{st}^{(0|2)}$, its
supersymmetry variation is $d$-exact.

Some final remarks are in order. In the previous section, we have seen the r%
\^{o}le of the PCO's of type $Z$ to reduce the picture of a given integral
form. Here we would like to apply the same technique to the $N=2$ case.

We start with the simplest volume form:
\begin{equation}
\omega ^{(1|2)}=V\delta (\psi )\delta (\bar{\psi})\,.  \label{NA}
\end{equation}%
It is an integral form, it is closed since it belongs to $\Omega ^{(1|2)}$,
but is also exact since it can be expressed as the differential of a $(0|2)$%
-form:
\begin{equation}
\omega ^{(1|2)}=d\left[ \frac{1}{2}V(\theta \iota +\bar{\theta}\bar{\iota}%
)\delta (\psi )\delta (\bar{\psi})\right]   \label{NB}
\end{equation}%
where the sign in the square bracket is opposite w.r.t. the sign of the PCO
in (\ref{SQMR}).\footnote{%
both terms in the r.h.s are necessary for $\omega $ to be real.} One can
verify that the integral of $\omega ^{(1|2)}$ on the supermanifold $\mathcal{%
M}^{(1|2)}$ vanishes. To avoid this problem, we need to construct a
different volume form and the easiest is
\begin{equation}
\mathrm{Vol}^{(1|2)}=V\theta \bar{\theta}\delta (\psi )\delta (\bar{\psi})\,.
\label{NC}
\end{equation}%
which is closed, but it is not exact. It is not manifestly supersymmetric,
but its supersymmetry variation is $d$-exact. Let us now apply a PCO of type
$Z$ to decrease the picture of the volume form:
\begin{align}
Z_{\psi }\left( \mathrm{Vol}^{(1|2)}\right) & \equiv \left[ d,\Theta (\iota
_{D})\right] \left( V\theta \bar{\theta}\delta (\psi )\delta (\bar{\psi}%
)\right) =d\left[ \Theta (\iota _{D})\left( V\theta \bar{\theta}\delta (\psi
)\delta (\bar{\psi})\right) \right]   \nonumber  \label{ND} \\
& =d\left[ V\theta \bar{\theta}\frac{1}{\psi }\delta (\bar{\psi})\right] =%
\left[ 2i\psi \bar{\psi}\theta \bar{\theta}\frac{1}{\psi }\delta (\bar{\psi}%
)-V\bar{\theta}\delta (\bar{\psi})\right] =-V\bar{\theta}\delta (\bar{\psi}%
)\,.
\end{align}%
where $\iota _{D}$ is the contraction operator along the odd vector field $D=%
\frac{\partial }{\partial \theta }+i\bar{\theta}\partial _{x}$. The
resulting pseudoform is closed, not exact, and it belongs to the space $%
\Omega ^{(1|1)}$. Then, we act with a PCO denoted by $Z_{\bar{\psi}}$
(putting now $D=\frac{\partial }{\partial \bar{\theta}}$) on the result of (%
\ref{ND}) and we obtain:
\begin{align}
Z_{\bar{\psi}}Z_{\psi }\left( \mathrm{Vol}^{(1|2)}\right) & =Z_{\bar{\psi}%
}\left( -V\bar{\theta}\delta (\bar{\psi})\right) =d\left[ \Theta (\iota _{%
\bar{D}})\left( -V\bar{\theta}\delta (\bar{\psi})\right) \right]   \nonumber
\label{NE} \\
& =d\left[ -V\bar{\theta}\frac{1}{\bar{\psi}}\right] =-2i\psi \bar{\psi}\bar{%
\theta}\frac{1}{\bar{\psi}}+V\bar{\psi}\frac{1}{\bar{\psi}}=V-2i\psi \bar{%
\theta}=\widetilde{V}^{(1|0)}
\end{align}%
where $\widetilde{V}^{(1|0)}$ is a closed superform in $\Omega ^{(1|0)}$. We
can finally check that this superform is the dual of the PCO $\mathbb{Y}%
^{(0|2)}$:
\begin{align}
\widetilde{V}^{(1|0)}\wedge \mathbb{Y}^{(0|2)}& =\left( V-2i\psi \bar{\theta}%
\right) \wedge \left[ -\frac{1}{2}iV\wedge (\theta \iota -\bar{\theta}\bar{%
\iota})\delta (\psi )\delta (\bar{\psi})\right]   \nonumber  \label{NF} \\
& =V\theta \bar{\theta}\delta (\psi )\delta (\bar{\psi})=\mathrm{Vol}%
^{(1|2)}\,.
\end{align}

\subsection{Coupling to gauge fields.}

In the case on $N=2$ theory, there is an additional symmetry ($R$-symmetry)
rotating the fermions
\begin{equation}
W^{\prime }=e^{i\alpha }W\,,~~~~\overline{W}^{\prime }=e^{-i\alpha }%
\overline{W}\,,  \label{GA_A}
\end{equation}%
and the gravitinos
\begin{equation}
\psi ^{\prime }=e^{i\alpha }\psi \,,~~~~\bar{\psi}^{\prime }=e^{-i\alpha }%
\bar{\psi}\,,  \label{GA_B}
\end{equation}%
To gauge this symmetry, we replace the differential $d$ in the action $%
\mathcal{L}_{N=2}^{(1|0)}$ by the covariant differential $\nabla $ such that
\begin{equation}
\nabla W=dW+iAW\,,~~~~~\nabla \overline{W}=d\overline{W}-iA\overline{W}%
\,,~~~~~  \label{GA_C}
\end{equation}%
The modifications appear only in the kinetic terms (in the present section,
we neglect the interaction terms) and we get
\begin{equation}
\frac{i}{2}\Big(\overline{W}\nabla W+\nabla \overline{W}W\Big)=\frac{i}{2}%
\Big(\overline{W}dW+d\overline{W}W\Big)-AW\overline{W}  \label{GA_D}
\end{equation}%
from which we derive the current $J=W\overline{W}V$ by taking the derivative
with respect the bosonic component of the gauge field.

We recall that a gauge field in the supermanifold $\mathcal{M}^{(1|2)}$ is
defined in terms of a $(1|0)$-connection
\begin{equation}
A=A_{x}V+A_{\psi}\psi+A_{\bar{\psi}}\bar{\psi}\,.   \label{GA_E}
\end{equation}
As always, the components of the $(1|0)$-connection exceed the physical
components, therefore we impose some constraints as follows. First, we
compute the field strength
\begin{align}
F=dA & =(DA_{x}-\partial_{x}A_{\psi})V\psi+(\bar{D}A_{x}-\partial_{x}\bar {A}%
_{\psi})V\bar{\psi}+DA_{\psi}\psi^{2}  \label{GA_F} \\
& +\Big(DA_{\bar{\psi}}+\bar{D}A_{\psi}+2iA_{x}\Big)\psi\bar{\psi}+\bar {D}%
A_{\bar{\psi}}\bar{\psi}^{2}\,.  \nonumber
\end{align}
Then we set
\begin{equation}
\Big(DA_{\bar{\psi}}+\bar{D}A_{\psi}+2iA_{x}\Big)=0\,,~~~~~~\bar{D}A_{\bar{%
\psi}}=0\,,~~~~~DA_{\psi}=0\,.   \label{GA_G}
\end{equation}
The first equation can be easily solved in terms of $A_{x}$ to get
\begin{equation}
A_{x}=\frac{i}{2}\Big(DA_{\bar{\psi}}+\bar{D}A_{\psi}\big)   \label{GA_H}
\end{equation}
with the condition that $A_{\psi}$ is anti-chiral and $A_{\bar{\psi}}$ is
chiral. As a consequence, by computing the combinations $DA_{x}-\partial
A_{\psi}$ and $(\bar{D}A_{x}-\partial A_{\bar{\psi}})$ we find that they
vanish. This implies that the full field strength $F$ vanishes. The
connection is given by
\begin{equation}
A=\frac{i}{2}\Big(DA_{\bar{\psi}}+\bar{D}A_{\psi}\Big)V+A_{\psi}\psi +A_{%
\bar{\psi}}\bar{\psi}\,.   \label{GA_I}
\end{equation}
The gauge symmetry is $\delta A_{\psi}=D\Lambda$ and $\delta A_{\bar{\psi}}=%
\bar{D}\Lambda$ where $\Lambda$ is the superfield gauge parameter. As a
consequence, $\delta A_{x}=\partial_{x}\Lambda$. Given the supermanifold $%
\mathcal{M}^{(1|2)}$, we can integrate the gauge field directly. For that we
need a $(1|2)$ integral form and therefore we multiply the gauge connection $%
A$ by the PCO's discussed above. We have
\begin{align}
A_{st}^{(1|2)} & =A^{(1|0)}\wedge{\mathbb{Y}}_{st}^{(0|2)}=\frac{i}{2}\Big(%
DA_{\bar{\psi}}+\bar{D}A_{\psi}\Big)V\theta\bar{\theta}\delta (\psi)\delta(%
\bar{\psi})\,,~~~~~  \label{GA_IA} \\
A^{(1|2)} & =A^{(1|0)}\wedge{\mathbb{Y}}^{(0|2)}=-\frac{i}{2}\Big(A_{\psi
}\psi+A_{\bar{\psi}}\bar{\psi}\Big)V(\theta\iota-\bar{\theta}\bar{\iota }%
)\delta(\psi)\delta(\bar{\psi})\,.  \nonumber
\end{align}
and one can show that:
\begin{equation}
\int_{\mathcal{M}^{(1|2)}}A_{st}^{(1|2)}=\int_{\mathcal{M}^{(1|2)}}A^{(1|2)}=%
\frac{i}{2}\int dx\left. (DA_{\bar{\psi}}+\bar{D}A_{\psi })\right\vert
_{\theta=\bar{\theta}=0}   \label{GA_IB}
\end{equation}
which is manifestly supersymmetric and gauge invariant if the supermanifold
has no boundary.

This approach can be followed to define a supersymmetric Wilson loop if
instead we choose a supermanifold $\mathcal{M}^{(1|2)}$ whose reduced
manifold is the circle $S^{1}$.

Notice that given a generic gauge connection $A^{(1|0)}$, there is no reason
for the two expressions $\int A_{st}^{(1|2)}$ and $\int A^{(1|2)}$ to match.
Indeed, as discussed above the choice of the PCO is arbitrary when the
superform $\mathcal{O}^{(1|0)}$ to which it is applied is $d$-closed.
Otherwise, it turns out that:
\begin{equation}
\mathcal{O}^{(1|0)}\wedge{\mathbb{Y}}_{st}^{(0|2)}=\mathcal{O}^{(1|0)}\wedge{%
\mathbb{Y}}^{(0|2)}+d\Big(\mathcal{O}^{(1|0)}\wedge\eta \Big)-d\mathcal{O}%
^{(1|0)}\wedge\eta   \label{GA_IC}
\end{equation}
where $\eta=iV\theta\bar{\theta}\iota\bar{\iota}\delta(\psi)\delta(\bar{\psi
})$ was computed in (\ref{SQMT}). Thus, if we integrate both members of (\ref%
{GA_IC}), the second term on r.h.s. drops out, but the third remains. If $%
\mathcal{O}^{(1|0)}$ is a connection form the above equation can be written
as:
\begin{equation}
{A}^{(1|0)}\wedge{\mathbb{Y}}_{st}^{(0|2)}={A}^{(1|0)}\wedge{\mathbb{Y}}%
^{(0|2)}+d\Big(A^{(1|0)}\wedge\eta\Big)-F^{(2|0)}\wedge\eta   \label{GA_ID}
\end{equation}
where we see that the last term vanishes if the field strength vanishes.

The Lagrangian is finally given by
\begin{equation}
\mathcal{L}_{N=2,kin~gauge}=\xi\Big(d\Phi-\psi\overline{W}+\bar{\psi }W\Big)-%
\frac{1}{2}(\xi^{2}+F^{2})V+\frac{i}{2}\Big(\overline{W}\nabla W+\nabla%
\overline{W}W\Big)   \label{GA_L}
\end{equation}
which is gauge invariant. The equations of motion are
\begin{align}
& d\Phi-\psi\overline{W}+\bar{\psi}W-\xi V=0\,,~~~~~  \label{GA_M} \\
& \nabla W=-\xi\psi\,,~~~~~ \\
& \nabla\overline{W}=i\xi\bar{\psi}\,, \\
& d\xi=0\,, \\
& F=0\,.   \label{GA_MA}
\end{align}
To check consistency of the equations, we act with $\nabla$ on the r.h.s. on
the fermionic equations and we get
\begin{equation}
iFW=-id\xi\wedge\psi-i\xi\nabla\psi\,.   \label{GA_N}
\end{equation}
The r.h.s. vanishes because $d\xi=0$ and $\nabla\psi=0$. The second equation
is the generalisation of $d\psi=0$ to the gauged version. It follows from (%
\ref{GA_N}) that the field strength $F$ vanishes. This is consistent with
the derivation outlined above.

Finally, if we consider the expression $J_{0}=-W\overline{W}$ (the function
appearing in the current for the R-symmetry) and we compute its differential
we obtain:
\begin{equation}
dJ_{0}=-i\xi(\psi\overline{W}+W\bar{\psi})\,.   \label{GA_O}
\end{equation}
The expression for $J_{0}$ is given in terms of superfields and belongs to a
supermultiplet. The supersymmetry variations can be computed directly from (%
\ref{GA_O}).

\subsection{$D=1,N=2$ Curved.}

To conclude this section, we analyze the curved manifold case. We replace
the flat supervielbein $V,\psi,\bar{\psi}$ with the curved ones $%
E^{v},E^{\psi },E^{\bar{\psi}}$ and we require that they satisfy the
constraints
\[
dE^{V}=2iE^{\psi}\wedge E^{\bar{\psi}}\,,~~~~~dE^{\psi}=0\,,~~~~~dE^{\bar {%
\psi}}=0\,.
\]
To solve these constraints we have to expand the superviebeins on a basis
\begin{align}
E^{V} & =E_{x}^{V}V+E_{\theta}^{V}\psi+E_{\bar{\theta}}^{V}\bar{\psi }
\label{CURB} \\
E^{\psi} & =E_{x}^{\psi}V+E_{\theta}^{\psi}\psi+E_{\bar{\theta}}^{\psi}\bar{%
\psi}  \nonumber \\
E^{\bar{\psi}} & =E_{x}^{\bar{\psi}}V+E_{\theta}^{\bar{\psi}}\psi +E_{\bar{%
\theta}}^{\bar{\psi}}\bar{\psi}\,.  \nonumber
\end{align}
The various components can be cast into a supermatrix. If we insert them
into the constraints, we find the final result
\begin{equation}
E^{V}=E_{\theta}^{\psi}E_{\bar{\theta}}^{\bar{\psi}}V\,,~~~~E^{\psi}=E_{%
\theta}^{\psi}\psi+\bar{D}E_{\theta}^{\psi}V\,,~~~~~E^{\bar{\psi}}=E_{\bar{%
\theta}}^{\bar{\psi}}\bar{\psi}+DE_{\theta}^{\psi}V\,.   \label{CURC}
\end{equation}
that resembles the $N=1$ case. All the equations can be solved in terms of
the two superfields $E_{\theta}^{\psi}$ and $E_{\bar{\theta}}^{\bar{\psi}}$
, that are anti-chiral and chiral respectively:
\begin{equation}
DE_{\theta}^{\psi}=0\,,~~~~\bar{D}E_{\bar{\theta}}^{\bar{\psi}}=0\,.
\label{CURD}
\end{equation}
As in the $N=1$ case we have:
\begin{align}
E^{V}\wedge\delta(E^{\psi})\delta(E^{\bar{\psi}}) & =E_{\theta}^{\psi }E_{%
\bar{\theta}}^{\bar{\psi}}V\wedge\delta\Big(E_{\theta}^{\psi}\psi+\bar {D}E_{%
\bar{\theta}}^{\bar{\psi}}V\Big)\wedge\delta\Big(E_{\bar{\theta}}^{\bar{\psi}%
}\bar{\psi}+DE_{\theta}^{\psi}V\Big)  \label{CURE} \\
& =E_{\theta}^{\psi}E_{\bar{\theta}}^{\bar{\psi}}V\wedge\frac{1}{E_{\theta
}^{\psi}}\delta\Big(\psi+\frac{1}{E_{\theta}^{\psi}}\bar{D}E_{\theta}^{\psi
}V\Big)\wedge\frac{1}{E_{\bar{\theta}}^{\bar{\psi}}}\delta\Big(E_{\bar{%
\theta }}^{\bar{\psi}}\bar{\psi}+DE_{\theta}^{\psi}V\Big)  \nonumber \\
& =V\wedge\delta\Big(\psi+\frac{1}{E_{\theta}^{\psi}}\bar{D}E_{\theta}^{\psi
}V\Big)\wedge\delta\Big(E_{\bar{\theta}}^{\bar{\psi}}\bar{\psi}+DE_{\theta
}^{\psi}V\Big)  \nonumber \\
& =V\delta(\psi)\delta(\bar{\psi})  \nonumber
\end{align}
Therefore also in the $N=2$ case the volume form is not modified going from
a flat to a curved supermanifold.

Let us analyse the PCO ${\mathbb{Y}}^{(0|2)}=-\frac{1}{2}iV(\theta\iota -%
\bar{\theta}\bar{\iota})\delta(\psi)\delta(\bar{\psi})$. We propose the
following curved version:
\begin{equation}
{\mathbb{Y}}^{(0|2)}=-\frac{i}{2}E^{V}(F\iota-\overline{F}\bar{\iota}%
)\delta(E^{\psi})\delta(E^{\bar{\psi}})   \label{CURF}
\end{equation}
where $F$ and $\overline{F}$ are two scalar superfields. We impose that this
integral form is closed
\begin{equation}
d{\mathbb{Y}}^{(0|2)}=-\frac{i}{2}E^{V}(\nabla F-\bar{\nabla}\overline {F}%
)\delta(E^{\psi})\delta(E^{\bar{\psi}})=0   \label{CURG}
\end{equation}
which implies that $\nabla F=\bar{\nabla}\overline{F}$. To solve this
equation, we note that we can use the same procedure as for the volume form
in (\ref{CURE}) to get:
\begin{align}
{\mathbb{Y}}^{(0|2)} & =-\frac{i}{2}E_{\theta}^{\psi}E_{\bar{\theta}}^{\bar{%
\psi}}V\wedge\Big(\frac{F}{(E_{\theta}^{\psi})^{2}E_{\bar{\theta}}^{\bar{\psi%
}}}\iota\delta(\psi)\delta(\bar{\psi})-\frac{\overline{F}}{%
E_{\theta}^{\psi}(E_{\bar{\theta}}^{\bar{\psi}})^{2}}\delta(\psi)\bar{\iota }%
\delta(\bar{\psi})\Big)  \label{CURH} \\
& =-\frac{i}{2}V\Big(\frac{F}{E_{\theta}^{\psi}}\iota\delta(\psi)\delta (%
\bar{\psi})-\frac{\overline{F}}{E_{\bar{\theta}}^{\bar{\psi}}}\delta (\psi)%
\bar{\iota}\delta(\bar{\psi})\Big)  \nonumber
\end{align}
from which it follows that:
\begin{equation}
F=\theta E_{\theta}^{\psi}\,,~~~~~~~\overline{F}=\bar{\theta}E_{\bar{\theta}%
}^{\bar{\psi}}   \label{CURK}
\end{equation}
They satisfy the condition $DF=E_{\theta}^{\psi}$ and $\bar{D}\overline {F}%
=E_{\bar{\theta}}^{\bar{\psi}}$. Therefore the PCO in the curved case is the
same as in the flat case.

\section{Quantization.}

Quantization for these simple systems can be obtained very easily. We
consider here the case $N=2$ with no superpotential ($\mathcal{W}=0$) for
simplicity. Furthermore, we promote the superfields $\Phi^{I}$ to be the
components of a multiplet $I=1,\dots,n$ describing a map
\[
\Phi^{I}(x,\theta,\bar{\theta}):\mathcal{M}^{(1|2)}\longrightarrow \mathcal{M%
}^{(n)}
\]
of the supermanifold into a $n$-dimensional Riemannian manifold $\mathcal{M}%
^{(n)}$.

From the equations of motion (\ref{GA_M}-\ref{GA_MA}), we find that the
solution is given by the following zero modes
\begin{align}
& \Phi^{I}=\phi_{0}^{I}+ixp_{0}^{I}+\lambda_{0}^{I}\bar{\theta}+\bar{\lambda
}_{0}^{I}\theta\,,~~~~~W^{I}=\lambda_{0}^{I}+p_{0}^{I}\theta\,,~~~~
\nonumber \\
& \overline{W}^{I}=-\bar{\lambda}^{I}-p_{0}^{I}\bar{\lambda}%
_{0}\,,~~~~\xi^{I}=p_{0}^{I}\,,~~~~F^{I}=0\,.   \label{QA}
\end{align}
where the zero modes $\phi_{0}^{I},p_{0}^{I},\lambda_{0}^{I},\bar{\lambda}%
_{0}^{I}$ satisfy the commutation relations
\begin{equation}
\lbrack\phi_{0}^{I},p_{0}^{J}]=i\hbar\,\eta^{IJ}\,,~~~~~~\{\lambda_{0}^{I},%
\bar{\lambda}_{0}^{J}\}=\hbar\,\eta^{IJ}\,.   \label{QB}
\end{equation}
and the Hilbert space $\mathcal{H}$ is constructed as follows:
\[
p_{0}^{I}|0\rangle=0\,,~~~~~~\lambda_{0}^{I}|0\rangle=0\,,~~~~~{\forall
I=1,\dots,n}
\]
and a generic state is given by:
\begin{align}  \label{ciccio}
|\chi\rangle=\sum_{p=0}^{n}|\chi,p\rangle=\sum_{p=0}^{n}\chi_{\lbrack
I_{1}\dots I_{n}]}(\phi_{0})\bar{\lambda}_{0}^{I_{1}}\dots \bar{\lambda}%
_{0}^{I_{n}}|0\rangle
\end{align}
where the functions $\chi_{\lbrack I_{1}\dots I_{n}]}(\phi_{0})$ are $L^{2}(%
\mathcal{M}^{(n)})$-integrable functions. The indices $I_{1}\dots I_{n}$ are
anti-symmetrized because of the Grassman variables $\bar{\lambda }_{0}^{I}$.

Let us project the Maurer-Cartan equations (\ref{SQMF}-\ref{SQMG}) on the
ground state.
\begin{align}  \label{QC}
d \Phi^{I} |0\rangle & = (i V p^{I}_{0} - \lambda^{I}_{0} \bar\psi-
\bar\lambda^{I}_{0} \psi) \, |0\rangle= - \psi\bar\lambda^{I}_{0} \,
|0\rangle= \psi\overline W^{I} |0\rangle  \nonumber \\
d \overline{W}^{I} |0\rangle & = - \bar\psi p^{I}_{0} |0\rangle= 0\,,
\end{align}
(by consistency $d W^{I} |0\rangle= \psi p^{I}_{0} |0\rangle= 0$).

Let us consider now a differential form of $\Omega^{\bullet}(\mathcal{M}%
^{(n)})$, written in local coordinates, applied to the ground state $%
|0\rangle:$
\begin{align}
\omega^{(p)}|0\rangle & =\omega_{{I_{1}\dots I_{p}}}(\Phi)d\Phi^{I_{1}}%
\wedge\dots\wedge d\Phi^{I_{p}}|0\rangle  \nonumber  \label{QD} \\
& =\psi^{p}\omega_{{I_{1}\dots I_{p}}}(\phi_{0})\bar{\lambda}%
_{0}^{I_{1}}\dots\bar{\lambda}_{0}^{I_{p}}|0\rangle=\psi^{p}|\omega,p\rangle
\end{align}
We obtain a map between the exterior bundle $\Omega^{\bullet}(\mathcal{M}%
^{(n)})$ and the Hilbert space. The powers of the gravitinos ($\psi^{p}$)
parametrize each state at a given fermion number. The right hand side of (%
\ref{QD}) must be interpreted as a superform on $\mathcal{M}^{(1|2)}$. That
would be impossible in the case $\mathcal{M}^{(1)}\longrightarrow \mathcal{M}%
^{(n)}$ of a pure bosonic $1$-dimensional manifold because the pullback of
any differential form on $\mathcal{M}^{(n)}$ gives always a $1$-form on $%
\mathcal{M}^{(1)}.$

In particular, we have a map:
\[
\Omega^{(p)}(\mathcal{M}^{(n)})\longrightarrow\Omega^{(p|0)}(\mathcal{M}%
^{(1|2)})
\]

Let us compute the action of the differential $d$ on a $p$-form:
\begin{align}
d\omega^{(p)}|0\rangle & =\partial_{K}\omega_{{I_{1}\dots I_{p}}}(\Phi
)d\Phi^{K}\wedge d\Phi^{I_{1}}\wedge\dots\wedge d\Phi^{I_{p}}|0\rangle
\label{QE} \\
& =\psi^{p}\bar{\psi}\,\partial^{K}\omega_{{[KI_{2}\dots I_{p}]}}(\phi _{0})%
\bar{\lambda}_{0}^{I_{2}}\dots\bar{\lambda}_{0}^{I_{p}}|0\rangle  \nonumber
\\
&+\psi^{p+1}\partial_{\lbrack K}\omega_{{I_{1}\dots I_{p}]}}(\phi_{0})\bar{%
\lambda}_{0}^{K}\bar{\lambda}_{0}^{I_{1}}\dots\bar{\lambda}%
_{0}^{I_{p}}|0\rangle  \nonumber
\end{align}
The closure of $\omega^{p}$ gives the equations:
\begin{equation}
\partial^{K}\omega_{{[KI_{2}\dots I_{p}]}}(\phi_{0})=0\,,~~~~~\partial
_{\lbrack K}\omega_{{I_{1}\dots I_{p}]}}(\phi_{0})=0\,,
\end{equation}
(where $\partial_{K}=\partial/\partial\phi_{0}^{K}$) that imply also $%
\partial^{2}\omega_{\lbrack I_{1}\dots I_{p}]}=0$. Therefore, the states of
the present theory are represented by on-shell $p$-forms of $\mathcal{M}%
^{(p)}$.

The closure and the co-closure of the differential form implies that $%
\omega^{(p)}$ is an harmonic form. We refer to \cite{vafa} for further
comments on this point.

We conclude the present section by studying two operators that share some
characteristics with the PCO's and have an action on the Hilbert space. We
consider the following two $(0|n)$-pseudoforms
\begin{equation}
\mathcal{Y}^{(0|n)}=\prod_{I=1}^{n}W^{I}\delta(dW^{I})\,,~~~~~\overline {%
\mathcal{Y}}^{(0|n)}=\prod_{I=1}^{n}\overline{W}^{I}\delta(d\overline{W}%
^{I})\,,
\end{equation}
They are non trivial elements of the cohomology $H_{d}^{(0|n)}$. We analyze
them from the quantum point of view and we act with a single pseudoform $%
W^{I}\delta(dW^{I})$ on the generic state $|\chi,q\rangle$ as given in (\ref%
{ciccio})
\begin{align}
\mathcal{Y}^{(0|n)}|\chi,q\rangle &
=W_{0}^{I}\delta(dW_{0}^{I})|\chi,q\rangle=W_{0}^{I}\delta(p_{0}^{I}\psi)|%
\chi,q\rangle  \nonumber  \label{QPCOB} \\
& =\lambda_{0}^{I}\delta(p_{0}^{I}\psi)\Big[\chi_{I_{1}\dots I_{q}}(\phi
_{0})\bar{\lambda}_{0}^{I_{1}}\dots\bar{\lambda}_{0}^{I_{q}}\Big]|0\rangle
\nonumber \\
& =q\frac{1}{\psi}\delta(p_{0}^{I})\chi_{II_{2}\dots I_{q}}(\phi_{0})\bar{%
\lambda}_{0}^{I_{2}}\dots\bar{\lambda}_{0}^{I_{q}}\Big]|0\rangle
\end{align}
The action of $\delta(p_{0}^{I})$ on the wave function $\chi_{I_{1}\dots
I_{q}}(\phi_{0})$ is computed using the integral representation of the Dirac
delta function:
\[
\delta(p_{0}^{I})\chi_{I_{1}\dots I_{q}}(\phi_{0})=\int_{-\infty}^{\infty
}du\,e^{iup_{0}^{I}}\,\chi_{I_{1}\dots
I_{q}}(\phi_{0})=\int_{-\infty}^{\infty}du\,\chi_{I_{1}\dots
I_{q}}(\phi_{0}^{K}+\delta_{I}^{K}u)
\]
since the exponential operator $e^{iup_{0}^{I}}$ acts as a finite
translation on the coordinate $\phi_{0}^{I}$. If the expression is
integrable, we are left with a wave function with a variable less. The
operator $\delta(p_{0}^{I})$ projects the quantum state into a zero-momentum
state along the direction $I$. Notice the appearance of the inverse of $\psi$%
. Note also that acting with this operator on (\ref{QD}), the inverse of $%
\psi$ reduces the power of $\psi$ appearing in (\ref{QD}). In the same way,
acting with $\overline{\mathcal{Y}}^{(0|n)}$ increases the power of $\bar{%
\lambda}_{0}^{I}$ (namely the degree form).

\section{A note on Observables.}

Since we are dealing with a quantum mechanical system, we are interested to
study the observables of the theory.

Let us suppose that the observables are identified by means of a nilpotent
charge $Q$ anticommuting with the differential $d$. For example, one can
consider one of the two supercharges $Q$ or $\bar{Q}$ (associated to each
supercharge there is a unit of the R-charge discussed in sec. 5.1, and we
refer to that unit charge (positive for $Q$ and negative for $\bar{Q}$) as
ghost number, since usually the form number and this number are identified
in the literature (see \cite{vafa}) with the bigrading of the BRST complex .

So, given $\mathcal{O}_{1}^{(0)}$ a ghost number $1$ form, we have the
sequence of descent equations:
\begin{equation}
\Big[Q,\mathcal{O}_{1}^{(0)}\Big]=0\,,~~~~~\Big[Q,\mathcal{O}_{0}^{(1)}\Big]%
=d\mathcal{O}_{1}^{(0)}   \label{OOA}
\end{equation}
where $\mathcal{O}_{0}^{(1)}$ has zero ghost-number and $1$ form degree,
while $\mathcal{O}_{1}^{(0)}$ has ghost number $1$ and zero form degree. The
operators $\mathcal{O}_{1}^{(0)}$ and $\mathcal{O}_{0}^{(1)}$ are written in
term of the superfields $\Phi$ and $W$ of the supersymmetric model with $N=1$%
. We notice that the integral of $\mathcal{O}_{0}^{(1)}$
\begin{equation}
\mathcal{T}=\int_{\mathcal{M}^{(1|1)}}\mathcal{O}_{0}^{(1)}\wedge{\mathbb{Y}}%
^{(0|1)},   \label{OOB}
\end{equation}
is invariant under the action of $Q$. Here we have introduced the PCO ${%
\mathbb{Y}}^{(0|1)}$ to convert the observable $\mathcal{O}_{0}^{(1)}$ into
an integral form of type $(1|1)$. The observable $\mathcal{O}_{0}^{(1)}$ is
closed (using the descent equations), and therefore one can suitably change
the PCO to a different (but cohomologous) one. Consequently, by changing the
PCO by an exact term we have:
\begin{equation}
\mathcal{O}^{(1)}\wedge\Big({\mathbb{Y}}^{(0|1)}+d\eta\Big)=\mathcal{O}%
^{(1)}\wedge{\mathbb{Y}}^{(0|1)}+d(\mathcal{O}^{(1)}\wedge\eta)   \label{OOC}
\end{equation}
which shows that a redefinition of the PCO amounts to a shift by exact terms
of the observables and this drops out from the integral in (\ref{OOB}).
Acting with the PCO on the descent equations (\ref{OOA}) we have:
\begin{align}
& \Big[Q,\mathcal{O}_{1}^{(0)}\Big]\wedge{\mathbb{Y}}^{(0|1)}=\Big[Q,\Big(%
\mathcal{O}_{1}^{(0)}\wedge{\mathbb{Y}}^{(0|1)}\Big)\Big]=0\,,~~~~~
\label{OOD} \\
& \Big[Q,\Big(\mathcal{O}_{0}^{(1)}\wedge{\mathbb{Y}}^{(0|1)}\Big)\Big]=d%
\Big(\mathcal{O}_{1}^{(0)}\wedge{\mathbb{Y}}^{(0|1)}\Big)
\end{align}
where we assumed that $[Q,{\mathbb{Y}}^{(0|1)}]=0$ (which implies that the
PCO is supersymmetric invariant). Notice that if the PCO is shifted by a $d$%
-exact term we have
\begin{align}
\mathcal{O}_{1}^{(0)}\wedge({\mathbb{Y}}^{(0|1)}+d\eta) & =\mathcal{O}%
_{1}^{(0)}\wedge{\mathbb{Y}}^{(0|1)}+d\Big(\mathcal{O}_{1}^{(0)}\wedge \eta%
\Big)-(d\mathcal{O}_{1}^{(0)})\wedge\eta  \label{OODA} \\
& =\mathcal{O}_{1}^{(0)}\wedge{\mathbb{Y}}^{(0|1)}+d\Big(\mathcal{O}%
_{1}^{(0)}\wedge\eta\Big)-\Big[Q,\mathcal{O}_{0}^{(1)}\Big]\wedge \eta
\nonumber \\
& =\mathcal{O}_{1}^{(0)}\wedge{\mathbb{Y}}^{(0|1)}+d\Big(\mathcal{O}%
_{1}^{(0)}\wedge\eta\Big)-\Big[Q,\mathcal{O}_{0}^{(1)}\wedge\eta\Big]
\nonumber
\end{align}
which shows that the variation of the PCO results into a $d$-exact term plus
a $Q$-exact term.

Associated to the complex $0\longrightarrow\Omega_{1}^{(0)}\longrightarrow
\Omega_{0}^{(1)}\rightarrow0$, (here $\Omega_{q}^{(p)}$ denotes the space of
the observables with quantum numbers $p$ and $q$) we have the complex of
\textit{integral} observables:
\[
0\longrightarrow\Omega_{1}^{(0|1)}\equiv\Omega_{1}^{(0)}\wedge{\mathbb{Y}}%
^{(0|1)}\longrightarrow\Omega_{0}^{(1|2)}\equiv\Omega_{0}^{(1)}\wedge{%
\mathbb{Y}}^{(0|1)}\longrightarrow0
\]
The choice of the PCO allows us to choose the representation most useful to
compute the correlators. The choice of a non-supersymmetric one reduces the
observable to the component fields, otherwise the choice of a supersymmetric
PCO produces observables in superspace.



\end{document}